\documentclass[nofootinbib]{revtex4}
\usepackage{graphicx}
\usepackage{amsmath}
\usepackage{amssymb}
\usepackage{float}
\usepackage{color}

\begin{document}
\title{Compact and extended objects from self-interacting phantom fields}
\author{
Vladimir Dzhunushaliev
}
\email{v.dzhunushaliev@gmail.com}
\affiliation{
Dept. Theor. and Nucl. Phys., KazNU, Almaty, 050040, Kazakhstan
}
\affiliation{
IETP, Al-Farabi KazNU, Almaty, 050040, Kazakhstan
}

\author{Vladimir Folomeev}
\email{vfolomeev@mail.ru}
\affiliation{Institute of Physicotechnical Problems and Material Science of the NAS
of the Kyrgyz Republic, 265 a, Chui Street, Bishkek 720071,  Kyrgyz Republic}

\author{Arislan Makhmudov}
\email{arslan.biz@gmail.com}
\affiliation{Peoples Friendship University of Russia, Educational and Scientific Institute of Gravitation and Cosmology}

\author{
Ainur Urazalina
}
\email{Y.A.A.707@mail.ru}
\affiliation{
Department of Physics, CSU Fresno, Fresno, CA, USA
}
\affiliation{
Dept. Theor. and Nucl. Phys., KazNU, Almaty, 050040, Kazakhstan
}

\author{
Douglas Singleton
}
\email{dougs@csufresno.edu}
\affiliation{Department of Physics, CSU Fresno, Fresno, CA 93740 USA
and \\
ICTP South American Institute for Fundamental Research,
UNESP - Univ. Estadual Paulista
Rua Dr. Bento T. Ferraz 271, 01140-070, S{\~a}o Paulo, SP, Brasi}

\author{
John Scott
}
\email{johnmlscott@mail.fresnostate.edu}
\affiliation{
Department of Physics, CSU Fresno, Fresno, CA, USA
}

\date{\today}

\begin{abstract}
In this work we investigate localized and extended objects for gravitating, self-interacting phantom fields. The
phantom fields come from two scalar fields with a ``wrong sign" (negative) kinetic energy term in the Lagrangian. This study covers
several solutions supported by these phantom fields: phantom balls, traversable wormholes, phantom cosmic strings, and 
``phantom" domain walls. These four systems are solved numerically and we try to draw out general, interesting features 
in each case. 
\end{abstract}
\maketitle

\section{Introduction}

The current astronomical and cosmological observations indicate that the Universe is in an epoch of accelerated expansion.
The source of this acceleration, dubbed dark energy (DE), is now under active investigation (for a review see the book~\cite{AmenTsu2010}
and references therein). One of the distinctive properties of DE is its large negative pressure, which is comparable in magnitude with its energy density.
To model DE, various approaches are employed,  from the simplest $\Lambda$CDM model, where the DE energy density comes
from a Cosmological Constant which does not change during the evolution of the Universe, to modified and multidimensional theories of gravity.

One of the most popular ways to describe DE is through models with various scalar fields.
For such fields one can introduce some effective equation of state $p=w \varepsilon$, relating the pressure of the fields 
({\it i.e.} $p$) with the energy density of the fields ({\it i.e.} $\varepsilon$). It then becomes possible to subdivide all such fields into 
two large classes: non-phantom fields, for which  $w>-1$ (for instance, quintessence scalar fields), and phantom fields, for which
the equation-of-state parameter $w$ is less than $-1$. The recent observational estimates give $w=-1.10\pm 0.14 (1\sigma)$
\cite{Komatsu:2010fb} and $w=-1.069_{-0.092}^{+0.091}$ \cite{Sullivan:2011kv}, {\it i.e.}, the possibility is not excluded that some of 
the matter filling the Universe acts as phantom matter.

When describing the accelerated expansion of the Universe, it is usually assumed
that DE is distributed homogeneously and isotropically on the largest scales.
This, however, does not exclude a possibility that DE may cluster on relatively small
scales comparable to sizes of galaxies or even of stars where one might expect an interesting interplay between
the normal tendency of gravity to pull matter together versus the tendency of DE to act repulsively.
For these cases one could form compact objects, dubbed DE stars,  both with a trivial and with a nontrivial topology
of the associated spacetime \cite{de_stars}. In particular, if these types of objects contain
phantom scalar fields one can  construct localized solutions that can be used in describing configurations both with a trivial
(star-like) \cite{Dzhunushaliev:2008bq} and with a nontrivial (wormhole-like) spacetime topologies~\cite{whs}.

Apart from the aforementioned star- and wormhole-like spherically symmetric configurations, one can consider extended objects
supported by scalar fields but having symmetries other than spherical.
One can look for cylindrically symmetric cosmic strings. Cosmic strings are extended configurations which could form
in the early Universe at phase transitions associated with spontaneous symmetry breaking \cite{Vilen}. Different types
of cosmic strings have been considered in the literature (see, {\it e.g.}, Ref.~\cite{Bran2}). Of particular interest to the phantom 
cosmic strings found in this paper are cosmic strings constructed from various types of scalar fields
\cite{Baze,Sant}, including models with two interacting scalar fields \cite{BezerradeMello:2003ei}.

Another category of extended objects are planar symmetric domain walls which are topological defects arising in 
both particle physics and cosmology (see, {\it e.g.}, Refs.~\cite{Bran2,Cvet} and references therein). They separate a spacetime
into several domains along a single coordinate.  The domain walls we find here arise from regions with fast spatial variation of
the scalar fields. Depending on the behavior of the field near the wall,  solutions describing such objects
can be subdivided into thin-wall solutions (see, e.g.,  \cite{thin_dom}), for which
the scalar-field energy density can be replaced by the delta function, and thick-wall solutions \cite{thick_dom}.

It is clear that the characteristics of the aforementioned compact and extended objects will depend on the specific type of fields employed
in their modeling. In the present paper we consider all four types of configurations (domain walls, boson ``stars", wormholes, cosmic strings)
constructed from two interacting phantom scalar fields. One caveat in regard to the ``star" solutions of the present work is that they
have masses on the order of the Planck mass -- they are not objects with a stellar mass. For this reason, the spherically symmetric solutions that we
find in this work are called phantom balls. The Systems with two ordinary scalar fields are well known from quantum field theory,
where they are used to obtain solitary wave solutions \cite{rajaraman}. When a gravitational field is present, such
systems have also been repeatedly considered in the cosmological and astrophysical contexts (see Ref.~\cite{2_fields_syst}).
In our previous works we have obtained a number of solutions with two scalar fields
(both normal scalar fields and phantom scalar fields) which can be employed to describe astrophysical objects and also
when considering cosmological solutions. We have considered regular spherically and cylindrically symmetric solutions
\cite{2_fields_our,Dzhunushaliev:2007cs,Dzhunushaliev:2015sla}; cosmological solutions~\cite{Dzhunushaliev:2006xh,Folomeev:2007uw};
thick brane solutions supported by normal and ghost scalar fields~\cite{2_fields_brane}. Here we will
extend those studies.

\section{General equations}
\label{GE}
In the simplest case a phantom scalar field can be introduced by changing the sign of the kinetic term in the Lagrangian.
Consistent with this, we choose the Lagrangian for two scalar fields $\phi,\chi$ as in Ref. \cite{Dzhunushaliev:2007cs}:
\begin{equation}
\label{lagrangian}
  L =-\frac{R}{16\pi G} - \left[
      \frac{1}{2}\partial_\mu \phi \partial^\mu \phi + \frac{1}{2}\partial_\mu \chi \partial^\mu \chi - V(\phi,\chi)
    \right]~,
\end{equation}
where $R$ is the scalar curvature, $G$ is the Newtonian gravitational constant. We have designated the 
two scalar fields  in \eqref{lagrangian} as phantom fields due to fact that we have taken their
kinetic terms ({\it i.e.} $-\frac{1}{2}\partial_\mu \phi \partial^\mu \phi$ and 
$-\frac{1}{2}\partial_\mu \chi \partial^\mu \chi$) to have the ``wrong" sign ({\it i.e.}  a $-$ sign) rather than a 
the ``correct" sign ({\it i.e.} a $+$ sign). More precisely though 
phantom fields require that the equation of state parameter $w$ \footnote{The equation of state parameter
is the ratio of the pressure to the energy density of the field/fluid namely $w \equiv \frac{p}{\rho}$} 
should satisfy $w<-1$. Putting a $-$ in front of the scalar field kinetic energy term can give $w<-1$, as in the original 
work on phantom fields by Caldwell \cite{caldwell}, but this is not necessarily true. We will find that for
most of our systems that our scalar fields
are phantom ({\it i.e.} $w<-1$) for some range of coordinates, but will be non-phantom ({\it i.e.} $w>-1$) for other
ranges of coordinates. The corresponding energy-momentum tensor will then be
\begin{equation}
\label{emt}
    T_{\mu}^\nu = - \left\{
        \partial_\mu \phi \partial^\nu \phi+
        \partial_\mu \chi \partial^\nu \chi-
        \delta_{\mu}^\nu \left[
            \frac{1}{2}\partial_\mu \phi \partial^\mu
            \phi + \frac{1}{2}\partial_\mu \chi \partial^\mu \chi - V(\phi,\chi)
        \right]
    \right\}~.
\end{equation}
Since $T^0 _0 =\rho$ and $T^i _i = - p$ we can use \eqref{emt} to calculate the equation of state parameter
$w =\frac{p}{\rho} = - \frac{T^i_i}{T^0 _0}$. As mentioned above we will find (for three of the four systems studied)
that the $w<-1$ for some range of spatial coordinates while being $w>-1$ for other ranges. Thus the scalar fields, 
$\phi , \chi$ are partially phantom fields in that they satisfy $w<-1$ for some range of coordinates, but do not give 
$w<-1$ everywhere. and variation of the Lagrangian \eqref{lagrangian} gives the Einstein and scalar field equations in the form
\begin{eqnarray}
\label{Einstein-gen}
    G_{\mu}^\nu &=&  \kappa T_{\mu}^\nu ~,
\\
\label{field-gen}
  \frac{1}{\sqrt{-g}}\frac{\partial}{\partial  x^\mu} \left[
    \sqrt{-g} g^{\mu\nu} \frac{\partial (\phi,\chi)}{\partial x^\nu}
  \right] &=& -\frac{\partial V}{\partial (\phi,\chi)} ~,
\end{eqnarray}
where $\kappa = 8 \pi G$. Below we will consider only static problems, for which
Eqs.~\eqref{Einstein-gen}-\eqref{field-gen} give the system of ordinary nonlinear differential equations with the potential energy
\begin{equation}
  \label{pot}
  V(\phi,\chi)=\frac{\lambda_1}{4}(\phi^2-m_1^2)^2+\frac{\lambda_2}{4}(\chi^2-m_2^2)^2+\phi^2
  \chi^2-V_0.
\end{equation}
Here  $m_1$ and $m_2$ are the  masses of the scalar fields, $\lambda_1, \lambda_2$
are the coupling constants, and $V_0$ is some constant whose value will be given for each of
the four types of solutions considered below. Our previous investigations of such kind of systems
\cite{2_fields_our,Dzhunushaliev:2007cs,Dzhunushaliev:2015sla,Dzhunushaliev:2006xh,Folomeev:2007uw,2_fields_brane}
indicate that regular solutions of the system \eqref{Einstein-gen}-\eqref{field-gen} with the potential
\eqref{pot} do exist only for certain eigenvalues of the parameters $m_1, m_2$. We will find the same result
here -- that only certain values of  $m_1, m_2$ lead to regular solutions.

In the following four sections we will present the details of the four different kind of phantom field solutions
-- planar ``phantom" domain wall, spherically symmetric phantom ball, phantom wormhole and cylindrical phantom cosmic string. Many
of the comments as to the physical viability of these various solutions (or not) we will save for the concluding section. 

\section{``Phantom" domain walls}
\label{DW}

We begin by studying the simplest extended solution supported by two phantom scalar fields -- the domain wall. 
The reason for the scare quotes around ``phantom" in the section heading is due to the fact that, although our
two scalar fields have the wrong sign in front of their kinetic energy terms, the equation 
of state parameter for these domain walls will never be less then $-1$ as we show at the end of this section. 
Since domain walls have planar symmetry we choose the metric in the form:
\begin{equation}
\label{metric_wall}
  ds^2=a^2(x) (dt^2-dy^2-dz^2)-dx^2.
\end{equation}
The metric function $a(x)$ depends on the coordinate $x$ only. This metric describes a 2D domain
wall embedded in a (3+1)-dimensional spacetime. Using Eqs.~\eqref{emt}, \eqref{Einstein-gen}, and \eqref{field-gen}, one can derive
the following Einstein and scalar field equations:
\begin{eqnarray}
\label{ein_wall_1}
  3\left(\frac{a^\prime}{a}\right)^2 &=& -\frac{1}{2}\left(
  \phi^{\prime 2} + \chi^{\prime 2}\right) + V ~,
\\
  \frac{a^{\prime \prime}}{a} - \left(\frac{a^\prime}{a}\right)^2 &=& \frac{1}{2}\left(\phi^{\prime 2} + \chi^{\prime 2}\right) ~,
\label{ein_wall_2} \\
\label{field_wall_1}
    \phi^{\prime \prime} + 3 \frac{a^\prime}{a}\phi^\prime &=& \phi\left[2\chi^2 + \lambda_1(\phi^2 - m_1^2)\right] ~,
\\
    \chi^{\prime \prime} + 3 \frac{a^\prime}{a}\chi^\prime &=&
    \chi\left[2\phi^2 + \lambda_2(\chi^2-m_2^2)\right] ~.
    \label{field_wall_2}
\end{eqnarray}
Here we have absorbed $\kappa$ via the following rescaling: $x/\sqrt \kappa \rightarrow x$,
$\phi \sqrt \kappa \rightarrow \phi$, $\chi \sqrt \kappa \rightarrow \chi$, and
$m_{1,2} \sqrt \kappa \rightarrow m_{1,2}$. The primes denote differentiation with respect to $x$. Looking at
\eqref{ein_wall_1} and \eqref{ein_wall_2} one can see that these are essentially the Friedman equations but with the
role of space and time coordinates exchanged ({\it i.e.} $t \leftrightarrow x$). The constant from the potential \eqref{pot} is chosen to be
\begin{equation}
\label{V0_wall}
  V_0=\frac{\lambda_1}{4} \left[ \phi^2 (0) -m_1^2 \right] ^2 + 
  \frac{\lambda_2}{4} \left[ \chi^2 (0) -m_2^2 \right] ^2+\phi ^2 (0)  \chi ^2 (0)  ~,
\end{equation}
where the initial values, $\phi (0)$ and $\chi (0)$, are given below. From \eqref{ein_wall_1} one can see that this choice for
$V_0$ ensures that $a^\prime (0) = 0$. From \eqref{emt} and for the metric in \eqref{metric_wall} the energy density is
\begin{equation}
  T^0_0 = - \frac{{\phi^\prime}^2 + {\chi^\prime}^2}{2} - V(\phi, \chi).
\label{energyDens-dw}
\end{equation}

We choose the boundary conditions at $x=0$ in the form:
\begin{alignat}{2}
    \label{ini1_wall}
    \phi(0)& = \sqrt{3},       & \qquad \phi^\prime(0)  &=0,
\nonumber \\
    \chi(0)&= 0.4, 0.6, 0.8, 1.1, 1.3
    & \qquad \chi^\prime(0)       &=0, \\
    a(0)&=1.0.
\nonumber
\end{alignat}
Following the procedure for finding solutions of Refs.~\cite{Dzhunushaliev:2007cs,Dzhunushaliev:2006xh}, we found
the masses $m_{1,2}$ presented in Table \ref{eignvlsDW}.

\begin{table}[H]
\begin{center}
\begin{tabular}{ |c|c|c|c|c|c|}
  \hline
   $\chi (0)$ & 0.4                    & 0.6                    & 0.8                   & 1.1                   & 1.3
\\ \hline
   $m_1 $   & 2.143597496    & 2.406090154     & 2.58469019     & 2.660693093   & 2.614816239 
   \\ \hline
   $m_2 $   & 2.64382729      & 2.796605957     & 2.98694478     & 3.348871789   & 3.655616217
\\
  \hline
\end{tabular}
\end{center}
  \caption{The eigenvalues $m_{1,2}$ versus $\chi (0)$ with $\phi (0) =\sqrt{3}$ for domain walls. 
	The coupling constants $\lambda_1=0.1, \lambda_2=1$.}
  \label{eignvlsDW}
\end{table}

\begin{figure}[H]
	\begin{minipage}[ht]{.45\linewidth}
		\begin{center}
			\fbox{
				\includegraphics[width=.9\linewidth]{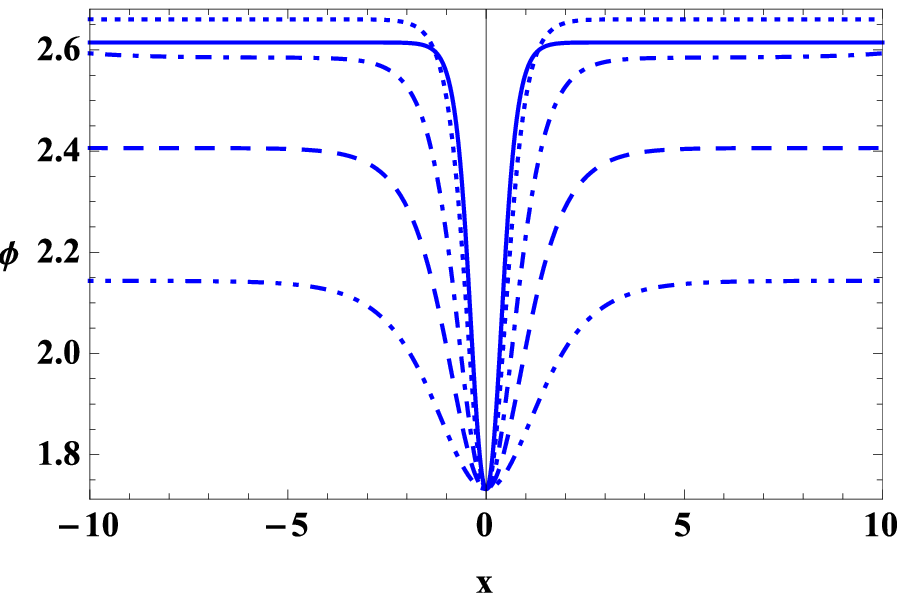}
			}
		\end{center}
		\caption{The scalar fields $\phi(x)$ for the domain walls.
             In Figs. \ref{phi_vs_x_wall} -- \ref{epsilon_vs_x_wall}
                    the solid curve corresponds to $\chi (0)  = 1.3$,
                    the dotted curve corresponds to $\chi (0)  = 1.1$,
                    the dashed - dotted corresponds to $\chi (0)  = 0.8$,
                    the dashed curve  corresponds to $\chi (0)  = 0.6$,
                    and dash - double dotted curve corresponds to $\chi (0)  = 0.4$.}
	\label{phi_vs_x_wall}
	\end{minipage}\hfill
	\begin{minipage}[ht]{.45\linewidth}
		\begin{center}
			\fbox{
				\includegraphics[width=.9\linewidth]{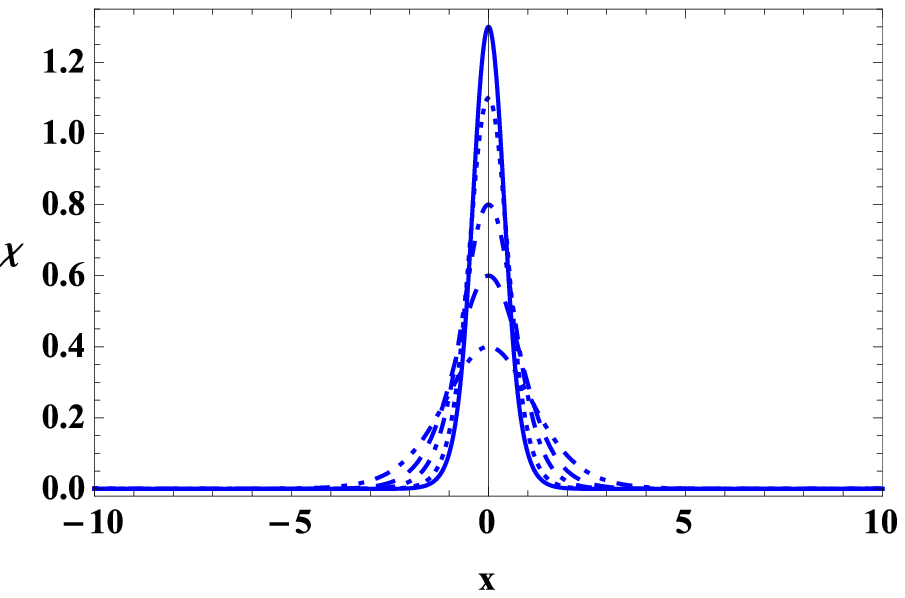}
			}
		\end{center}
		\caption{The scalar fields $\chi(x)$ for the domain walls.}
	\label{chi_vs_x_wall}
	\end{minipage}
	\begin{minipage}[ht]{.45\linewidth}
		\begin{center}
			\fbox{
				\includegraphics[width=.9\linewidth]{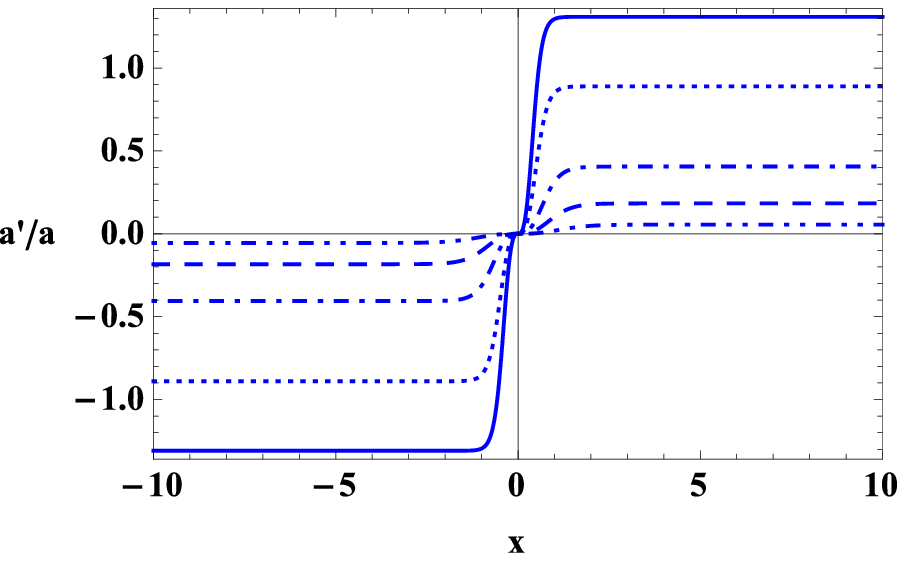}
			}
		\end{center}
		\caption{The functions $a'(x)/a(x)$ for the domain walls.}
	\label{a_vs_x_wall}
	\end{minipage}\hfill
	\begin{minipage}[ht]{.45\linewidth}
		\begin{center}
			\fbox{
				\includegraphics[width=.9\linewidth]{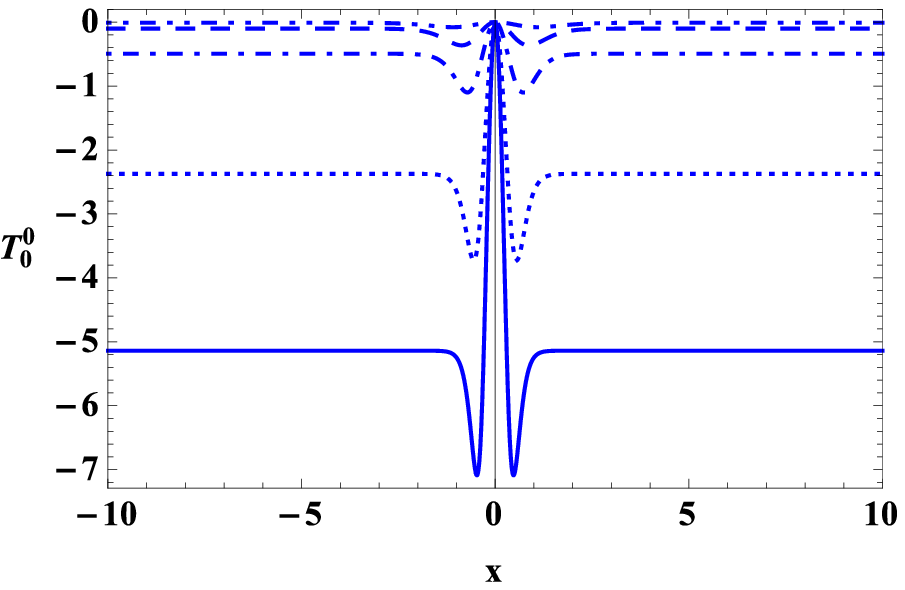}
			}
		\end{center}
		\caption{The energy densities $T^0_0(x)$ for the domain walls.}
	\label{epsilon_vs_x_wall}
	\end{minipage} \hfill
	\begin{minipage}[ht]{.45\linewidth}
		\begin{center}
			\fbox{
				\includegraphics[width=0.9\linewidth]{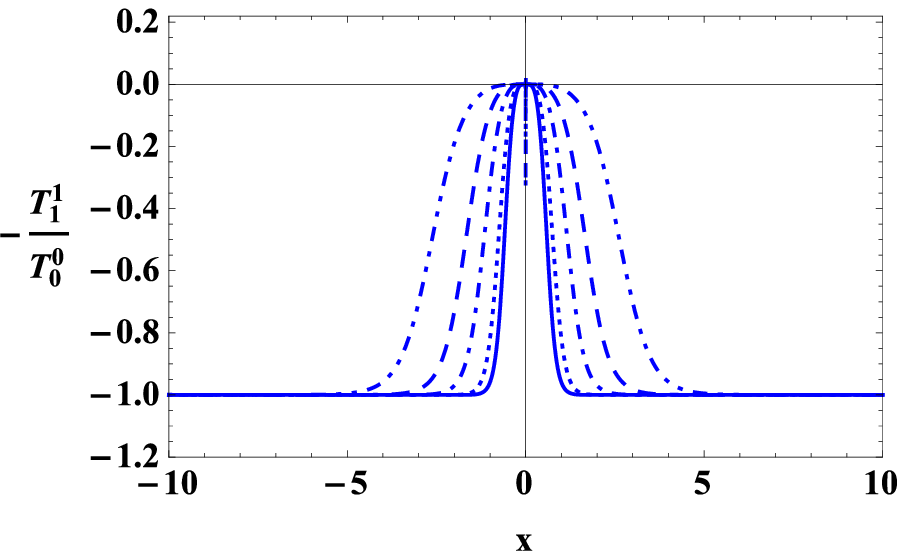}
			}
			\caption{The $x$-direction equation of state for the domain wall, $w = p_{x}(x)/\epsilon(x)$.}
			\label{F_wh_sine}
		\end{center}
	\end{minipage}\hfill 
\end{figure}

The results of numerical calculations for the  scalar fields are given in Figs.~\ref{phi_vs_x_wall}, \ref{chi_vs_x_wall};
for the metric functions $a(x)$ -- in Fig.~\ref{a_vs_x_wall}; and for the energy density $T_0^0$ -- in Fig.~\ref{epsilon_vs_x_wall}.

Let us estimate an asymptotic  behavior of the solutions. One can see from Eq.~\eqref{ein_wall_2} that the right-hand side goes
asymptotically to zero, and the solution of this equation is
\begin{equation}
\label{asym_a}
  a (x) \approx a_0 e^{\alpha x},
\end{equation}
where $a_0$ and $\alpha$ are integration constants. This solution corresponds to an anti-de Sitter-like solution for the spatial variable $x$.
Then, using \eqref{asym_a} we look for the asymptotic form of the scalar fields from Eqs.~\eqref{field_wall_1}-\eqref{field_wall_2}. Asymptotically 
we find that $\phi$ approaches $m_1$ and $\chi$ approaches zero so we write the field $\phi$ as $\phi \approx m_1 - \delta \phi$
where $\delta \phi \ll 1$ as $x \rightarrow \pm\infty$. In this way we obtain from Eqs.~\eqref{field_wall_1}-\eqref{field_wall_2} 
the asymptotic equations for $\delta \varphi$ and $\chi$:
\begin{eqnarray}
\label{as-varphi_wall}
  \delta \phi^{\prime \prime} + 3\alpha \delta \phi^\prime &\approx&
  2 \lambda_1 m_1^2 \delta \phi ,
\\
\label{as-chi_wall}
  \chi^{\prime \prime} + 3\alpha \chi^\prime &\approx&
  (2 m_1^2-\lambda_2 m_2^2) \chi.
\end{eqnarray}
The damping solutions to these equations are
\begin{eqnarray}
    \delta \phi &\approx&  C_{\phi} \exp{\left[-\frac{x}{2}\left(3\alpha+
    \sqrt{9\alpha^2+8\lambda_1 m_1^2}\right)\right]},
\label{sol1_wall} \\
    \chi &\approx&  C_{\chi}
    \exp{\left[-\frac{x}{2}\left(3\alpha+
    \sqrt{9\alpha^2+4(2 m_1^2-\lambda_2 m_2^2)}\right)\right]},
\label{sol2_wall}
\end{eqnarray}
where $C_{\phi}, C_{\chi}$ are integration constants. So we have solutions that tend asymptotically to the local minimum of the potential
\eqref{pot} at $\varphi=m_1$ and $\chi=0$. Note that with this asymptotic form of the scalar fields and for the choice of potential constant
$V_0$ from \eqref{V0_wall} asymptotically $T^0 _0 <0$ as seen in Fig. \ref{epsilon_vs_x_wall}. Actually $T^0 _0 <0$ for the entire range of
$x$. While potentially unphysical this negative energy density is not unexpected since in the case of some other types of planar solution 
\cite{jones-ajp} one also finds $T^0_0 <0$.

Finally in Fig. \eqref{F_wh_sine} we have given the equation of state parameter which is
the ratio of the pressure to energy density of the field namely $w=\frac{p_x}{\rho} = -\frac{T^1 _1}{T^0 _0}$
where $p_x = - T^1_1 = - \left( \frac{{\phi'}^2 + {\chi'}^2}{2} - V \right)$ and 
$\rho = T^0 _0 = - \frac{{\phi'}^2 + {\chi'}^2}{2} - V$. The subscript $x$ indicates this is the pressure in the $x$
direction. The different components of the scalar field energy-momentum tensor are
calculated via \eqref{emt}. There are also equation of state parameters for the two directions
perpendicular to $x$ which are determined via the pressures, 
$-T^2_2$ and $-T^3_3$. Since we only assumed $x$-dependence these two perpendicular
equation of state parameters are identically equal to $-1$ {\it i.e.} $-\frac{T^2 _2}{T^0 _0} = -\frac{T^3 _3}{T^0 _0} =-1$.
In contrast the radial equation of state parameter, $w= -\frac{T^1 _1}{T^0 _0}$, has more interesting behavior as shown
in Fig. \eqref{F_wh_sine} -- it starts at $w=0$ at $x=0$ and asymptotically approaches $-1$. Thus for this case
the scalar fields are not phantom fields since $w \ge -1$ for all $x$. Thus one can not call these phantom domain walls. This is
the reason for the scare quotes in the section heading.

\section{Phantom ball}
\label{SSS}

In looking for spherically symmetric solutions of the above system of gravity plus two phantom scalar fields, 
with the requirement of trivial topology, we find that unlike our previous study \cite{Dzhunushaliev:2008bq} 
there are no boson star solutions. We do find spherically symmetric solution but the ``mass" is 
of the order of the Planck mass rather than of astrophysical scale. Furthermore the mass of these solutions
(obtained by integrating their energy density) leads to negative masses,
making the physical use of these solutions extremely questionable. In our previous study we 
assumed spherically symmetric solutions supported by one (complex) scalar field. Here we consider such spherically 
symmetric configurations created by two real phantom scalar fields.

To do so, let us choose the metric in Schwarzschild coordinates:
\begin{equation}
\label{metric_sph}
  ds^2 = B(r)dt^2 - A(r)dr^2 - r^2(d\theta^2 + \sin^2\theta d\varphi^2).
\end{equation}
Then the Einstein equations~\eqref{Einstein-gen} and the scalar field equations~\eqref{field-gen} will be:
\begin{eqnarray}
\label{einst1_sph}
    \frac{1}{r}\frac{A^\prime}{A^2}+\frac{1}{r^2}\left(1-\frac{1}{A}\right)&=&
    -\frac{1}{2A}\left(\phi^{\prime 2}+\chi^{\prime 2}\right)-
    V(\phi,\chi),
\\
    \frac{1}{r}\frac{B^\prime}{A B}-\frac{1}{r^2}\left(1-\frac{1}{A}\right)&=&
    -\frac{1}{2A}\left(\phi^{\prime 2}+\chi^{\prime 2}\right)+V(\phi,\chi),
\label{einst2_sph}
\\
\frac{B^{\prime \prime}}{B}-\frac{1}{2}\left(\frac{B^\prime}{B}\right)^2-\frac{1}{2}\frac{A^\prime}{A}\frac{B^\prime}{B}
-\frac{1}{r}\left(\frac{A^\prime}{A}-\frac{B^\prime}{B}\right)&=& 2A\left[\frac{1}{2A}\left(\phi^{\prime 2}+\chi^{\prime 2}\right)+
    V(\phi,\chi)\right],
\label{einst3_sph} \\
\label{sfe1_sph}
    \phi^{\prime \prime}+\left(\frac{2}{r}+\frac{B^\prime}{2B}-\frac{A^\prime}{2A}\right)\phi^\prime &=&
    A\phi\left[2\chi^2+\lambda_1(\phi^2-m_1^2)\right]~,
\\
    \chi^{\prime \prime}+\left(\frac{2}{r}+\frac{B^\prime}{2B}-\frac{A^\prime}{2A}\right)\chi^\prime &=&
    A\chi\left[2\phi^2+\lambda_2(\chi^2-m_2^2)\right]~.
\label{sfe2_sph}
\end{eqnarray}
In order to numerically solve the above system of coupled equations we have introduced
dimensionless variables via the following definitions: $r/\sqrt \kappa \rightarrow r$,
$\phi \sqrt \kappa \rightarrow \phi$, $\chi \sqrt \kappa \rightarrow \chi$, and $m_{1,2} \sqrt \kappa \rightarrow m_{1,2}$. 
The primes denote differentiation with respect to the rescaled $r$. Since from the solutions below
we find that the asymptotic values of the scalar fields are $\phi = m_1$ and $\chi =0$ we pick
the constant in the potential \eqref{pot} as $V_0 = (1/4)\lambda_2 m_2^4$. This has results in
the energy density being equal zero as $r \rightarrow \infty$. Eq.~\eqref{einst3_sph} is a consequence of
Eqs.~\eqref{einst1_sph} and \eqref{einst2_sph} so we only have to solve the latter two equations for the metric
functions $A(r), B(r)$.

Using Eqs.~\eqref{einst1_sph}-\eqref{sfe2_sph}, one can find the following asymptotic behavior of the metric functions $A(r)$, 
$B(r)$ and phantom scalar fields $\phi, \chi$:
\begin{eqnarray}
\label{asym_sph}
    A &\approx& 1 - \frac{r_0}{r} ,
\\
\label{asym_sph2}
    B &\approx& B_\infty \left( 1 + \frac{r_0}{r} \right) ,
\\
  \phi &\approx&
  m_1 - C_{\phi} \frac{\exp{\left(- \sqrt{2 \lambda_1 m_1^2} \,\, r \right)}}{r} ,
\label{sol1b} \\
  \chi &\approx&
  C_{\chi}\frac{\exp{\left(- \sqrt{ (2 m_1^2-\lambda_2 m_2^2)} \,\,r\right)}}{r},
\label{sol2b}
\end{eqnarray}
where $r_0$ and $B_\infty$ are constants; $C_{\phi}, C_{\chi}$ are integration constants.
In effect, $r_0$ defines the total mass of the system and $B_\infty$ -- the rate of flow of time at infinity.
By rescaling the time variable $t$, we can provide $B_\infty=1$ as $r \rightarrow \infty$, {\it i.e.}, asymptotically
we have flat Minkowski spacetime.

Here we expand on the studies of the spherically symmetric solutions performed in Ref.~\cite{Dzhunushaliev:2007cs}.
For this purpose, we will consider the properties of solutions depending on the central value of the scalar field $\chi_0$.
The boundary conditions at $r=0$, for starting our numerical integration, are taken in the form:
\begin{alignat}{2}
    \label{ini1_sph}
    \phi(0)&=0.5,       & \qquad \phi^\prime(0)  &=0, \nonumber \\
    \chi(0)&= 0.2,0.5,0.8,1.0,1.2,
    & \qquad \chi^\prime(0)       &=0, \\
    A(0)&=1.0,&  \qquad B(0)&=1.0.\nonumber
\end{alignat}
Since \eqref{einst1_sph} and \eqref{einst2_sph} are first order in the derivative we only need $A(0) , B(0)$. Recall that
\eqref{einst3_sph} is redundant so in numerically solving the coupled system we only use
\eqref{einst1_sph} , \eqref{einst2_sph}, \eqref{sfe1_sph}, and \eqref{sfe2_sph}.
For a given value of $\chi (0)$ there are only certain values of $m_1, m_2$ which lead to regular solutions. These values of $m_1, m_2$
are found via the procedure described in \cite{Dzhunushaliev:2007cs,Dzhunushaliev:2006xh}.
The masses $m_{1,2}$ which yield regular solutions are shown in Table \ref{eignvls} as a function of $\chi(0)$. One may think of
$m_1, m_2$ as eigenvalues.

\begin{figure}[H]
	\begin{minipage}[ht]{.45\linewidth}
		\begin{center}
			\fbox{
				\includegraphics[width=.9\linewidth]{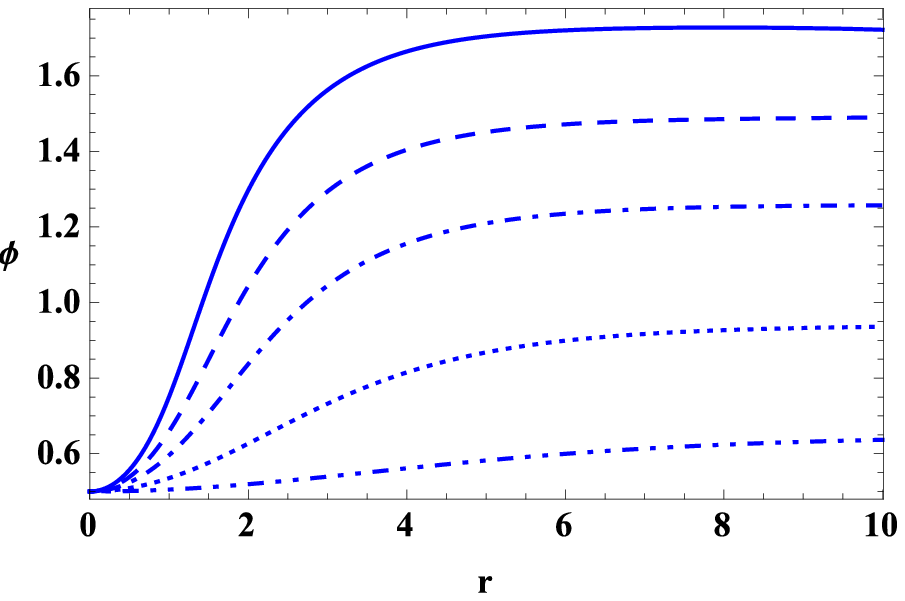}
			}
		\end{center}
		\caption{The scalar fields $\phi(r)$ for the phantom ball. In Figs. \ref{chi_vs_x_boson_stars} -- \ref{mass_vs_chi0_boson_stars}
                    the solid curve corresponds to $\chi (0)  = 1.2$,
                    the dashed curve  corresponds to $\chi (0)  = 1.0$,
                    the dashed - dotted corresponds to $\chi (0)  = 0.8$,
                    the dotted curve corresponds to $\chi (0)  = 0.5$,
                    and dash - double dotted curve corresponds to $\chi (0)  = 0.2$.
                    }
	\label{chi_vs_x_boson_stars}
	\end{minipage}\hfill
	\begin{minipage}[ht]{.45\linewidth}
		\begin{center}
			\fbox{
				\includegraphics[width=.9\linewidth]{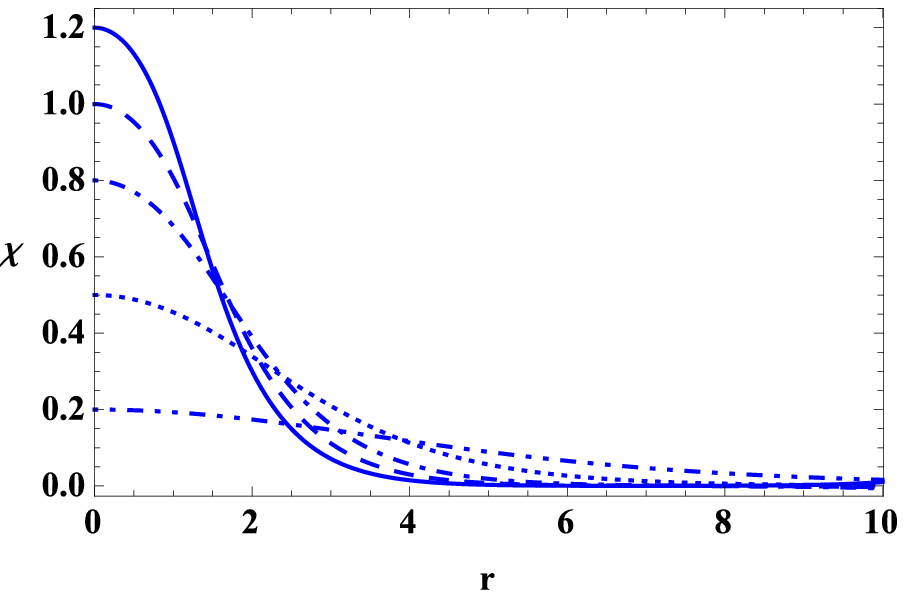}
			}
		\end{center}
		\caption{The scalar fields $\chi(r)$ for the phantom ball.
									}
	\label{phi_vs_x_boson_stars}
	\end{minipage}
	\begin{minipage}[ht]{.45\linewidth}
		\begin{center}
			\fbox{
				\includegraphics[width=.9\linewidth]{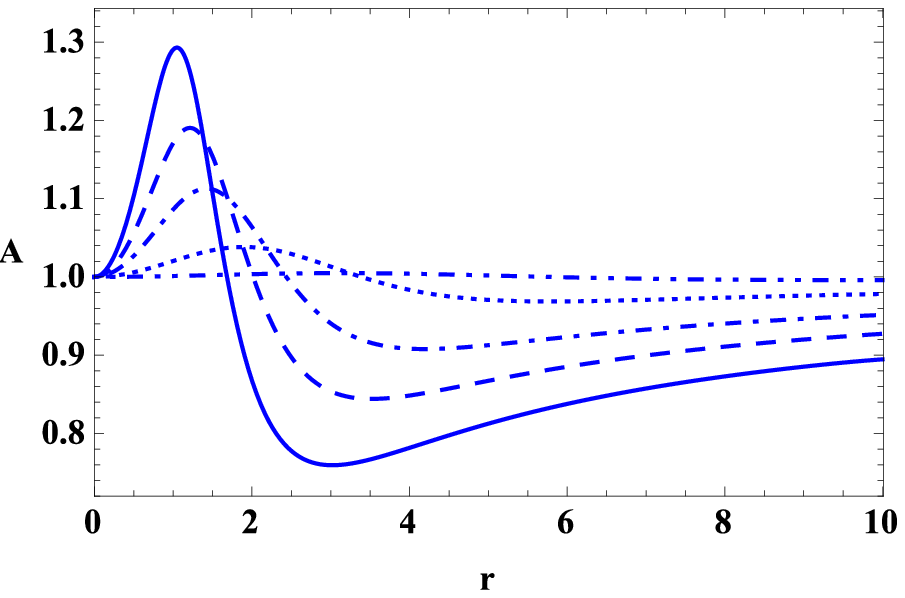}
			}
		\end{center}
		\caption{The metric functions $A(r)$ for the phantom ball.
									}
	\label{A_vs_x_boson_stars}
	\end{minipage}\hfill
	\begin{minipage}[ht]{.45\linewidth}
		\begin{center}
			\fbox{
				\includegraphics[width=.9\linewidth]{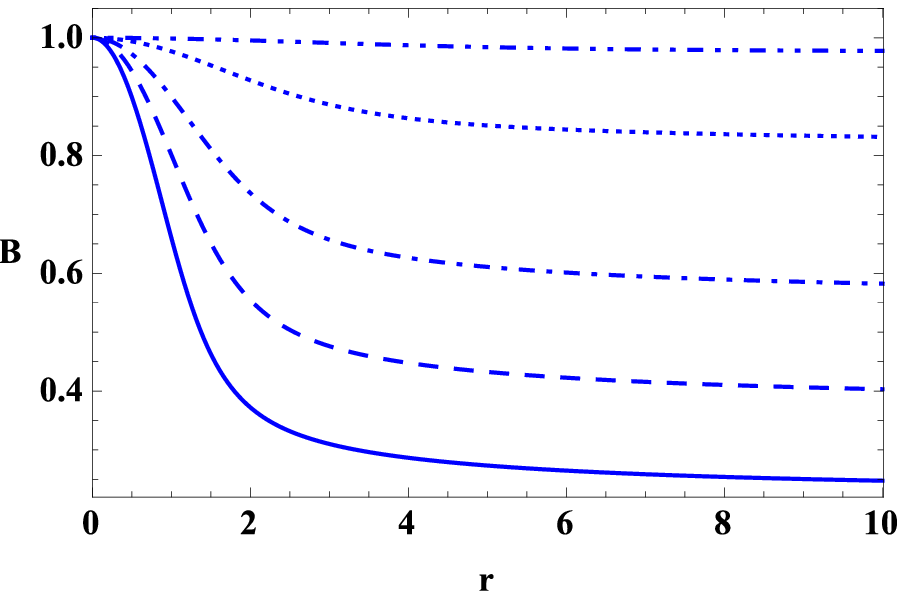}
			}
		\end{center}
		\caption{The metric functions $B(r)$ for the phantom ball.
									}
	\label{B_vs_x_boson_stars}
	\end{minipage}
	\begin{minipage}[ht]{.45\linewidth}
		\begin{center}
			\fbox{
				\includegraphics[width=.9\linewidth]{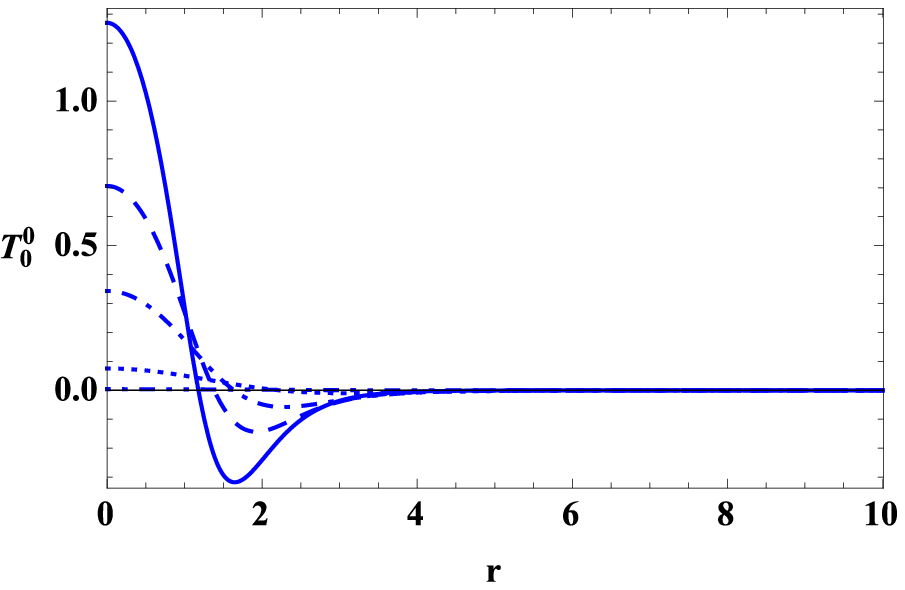}
			}
		\end{center}
		\caption{The energy density profiles for the phantom ball.
									}
	\label{epsilon_vs_x_boson_stars}
	\end{minipage}\hfill
		\begin{minipage}[ht]{.45\linewidth}
		\begin{center}
			\fbox{
				\includegraphics[width=0.9\linewidth]{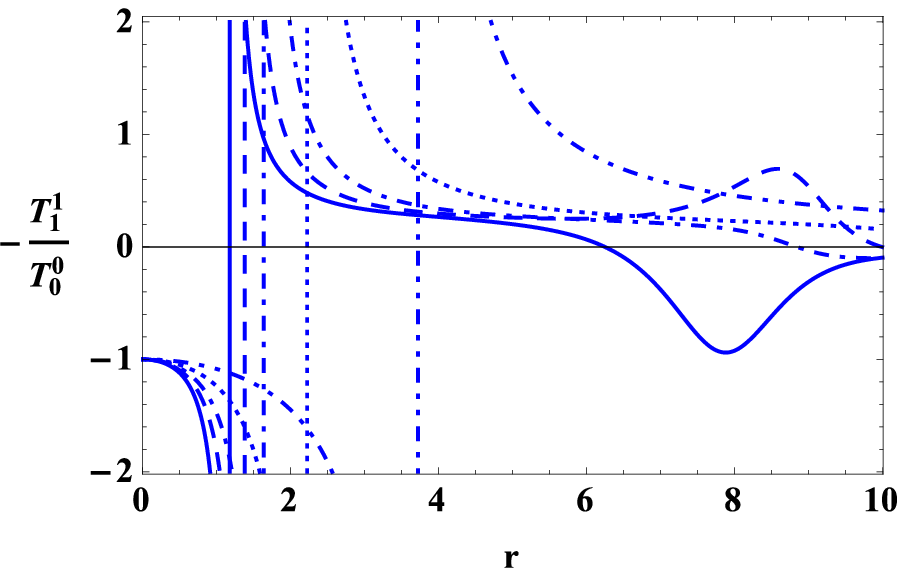}
			}
			\caption{The radial equation of state for the phantom ball $w = p_{r}(r)/\epsilon(r)$.}
			\label{F_wh_sine1}
		\end{center}
	\end{minipage}\hfill 
	\begin{minipage}[ht]{.45\linewidth}
		\begin{center}
			\fbox{
				\includegraphics[width=.9\linewidth]{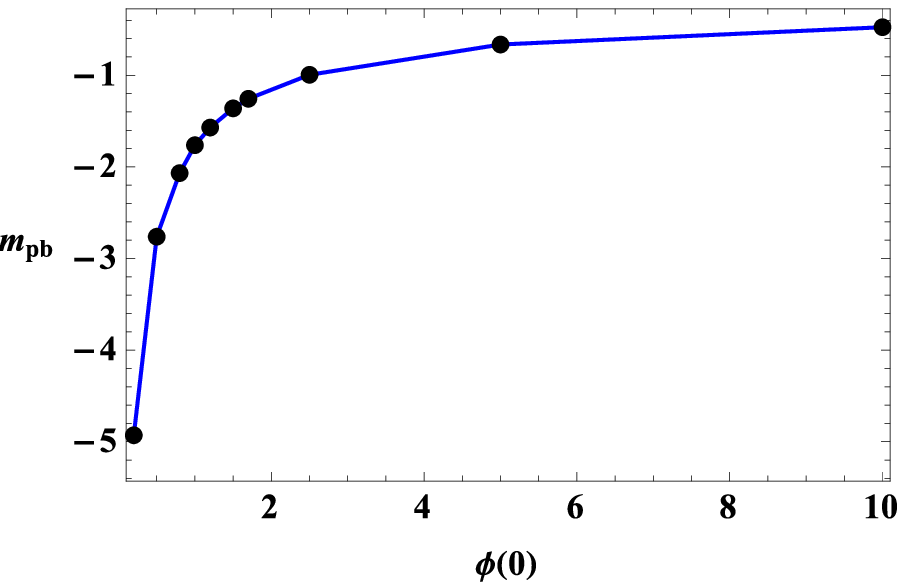}
			}
		\end{center}
		\caption{The mass of the phantom ball for $\chi (0) = 0.5$ and as a function 
		of $\phi (0) =  0.2,0.5,0.8,1.,1.2,1.5,1.7,2.5,5,10$. 
}
	\label{mass_vs_chi0_boson_stars}
	\end{minipage}
\end{figure}

The results of the numerical calculations for the scalar fields are presented in Figs.
\ref{chi_vs_x_boson_stars}, \ref{phi_vs_x_boson_stars};
for the metric functions $A(r), B(r)$ -- in Figs.~\ref{A_vs_x_boson_stars}, \ref{B_vs_x_boson_stars};
and for the energy density $T_0^0$ -- in Fig.~\ref{epsilon_vs_x_boson_stars}.

\begin{table}[H]
\begin{center}
\begin{tabular}{ |c|c|c|c|c|c|}
  \hline
   $\chi (0)$& 0.2           & 0.5             & 0.8           & 1.              & 1.2        \\ \hline
   $m_1$    & 0.64929   & 0.941025   & 1.26         & 1.4882      & 1.73298\\ \hline
   $m_2$    & 0.86731   & 1.14404     & 1.42583   & 1.61421    & 1.80206\\
  \hline
\end{tabular}
\end{center}
  \caption{The eigenvalues $m_{1,2}$ versus $\chi (0)$ with $\phi (0) = 0.5$ for phantom ball solutions.  The coupling constants are
	taken as $\lambda_1=0.1, \lambda_2=1$.}
  \label{eignvls}
\end{table}

In Fig. \eqref{F_wh_sine1} we have plotted the radial equation of state parameter,  $w=\frac{p_r}{\rho} = -\frac{T^1 _1}{T^0 _0}$,
for the phantom ball. As was the case with the domain wall, the equation of state parameters for the two perpendicular
directions ({\it i.e.} $w=\frac{p_\theta}{\rho} = -\frac{T^2 _2}{T^0 _0}$ and $w=\frac{p_\varphi}{\rho} = -\frac{T^3 _3}{T^0 _0}$)
are identically equal to $-1$. The divergence of $w$ seen in Fig. \eqref{F_wh_sine1} comes from the places where 
$T^0 _0 \rightarrow 0$ as can be seen in Fig. \eqref{epsilon_vs_x_boson_stars}.  

From the equation of state parameter in Fig. \eqref{F_wh_sine1} we find that for some range of $r$ near the 
origin the two scalar fields are phantom fields with $w<-1$, but for $r \rightarrow \infty$ the scalar fields 
have less exotic equations of state and for some parameters go to the dust equation of state $w=0$. 
It is because of the phantom behavior near $r=0$ that we call these solutions ``phantom ball" solutions. 
To further examine the physical character of these solutions we
turn to their total mass. We begin by defining the dimensionless mass of a phantom ball as
\begin{eqnarray}
  m_{pb} &=& 4 \pi \int \limits_0^\infty r^2 \sqrt{A(r)} \epsilon_{pb} dr ,
\label{massBS} \\
  \epsilon_{pb} &\equiv& T^0_0 =
  - \frac{{\phi^\prime}^2 + {\chi^\prime}^2}{2 A} - V(\phi, \chi),
\label{energyDens}
\end{eqnarray}
where $\epsilon_{pb}$ is the total energy density. This corresponds to the proper mass of the system \cite{wald} 
which is the Newtonian mass of the system ({\it i.e.} $4 \pi \int \limits_0^\infty r^2 \epsilon_{pb} dr$) 
plus the binding energy. The profile of $m_{pb}$ as a function of $\phi (0)$ for
$\chi (0) = 0.5$ is presented in Fig. \ref{mass_vs_chi0_boson_stars}. The values of $\phi(0)$ are 
$\phi (0) =  0.2,0.5,0.8,1.,1.2,1.5,1.7,2.5,5,10$. For each different $\phi(0)$ we needed to calculate 
the associated masses $m_1 , m_2$. The main thing to note is that $m_{pb} <0$ which makes the physical use of these
solutions questionable. Also, taking into account the rescaling that was performed on the coordinates, fields and
parameter, the magnitude of $m_{pb}$ is of the order of the Planck mass. Thus even if
$m_{pb}$ were positive these would not be astrophysical solution but would be a Planck mass particle like 
solution. One could compare the phantom ball solutions to GUT magnetic monopole solutions \cite{thooft}. Unlike GUT monopoles
the present solutions do not involve gauge fields. In the next section by changing our metric form 
from Schwarzschild-like to wormhole-like, we do find positive mass solutions for some parameters.  

\section{Phantom traversable wormholes}
\label{pwh}

\begin{figure}[H]
	\begin{minipage}[ht]{.45\linewidth}
		\begin{center}
			\fbox{
				\includegraphics[width=0.9\linewidth]{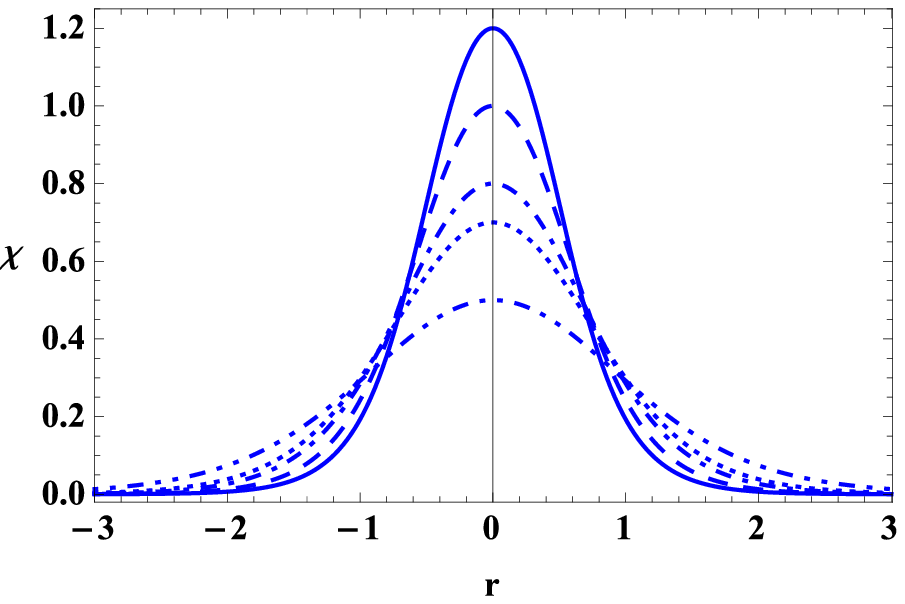}
			}
		\end{center}
		\caption{The behavior of the scalar field $\chi(r)$ for the traversable wormhole.
		In Figs. \ref{chi_vs_x} -- \ref{epsilon_vs_x} 
		the solid curve corresponds to $\chi (0)  = 1.2$,
                    the dashed curve  corresponds to $\chi (0)  = 1.0$,
                    the dashed - dotted corresponds to $\chi (0)  = 0.8$,
                    the dotted curve corresponds to $\chi (0)  = 0.7$,
                    and dash - double dotted curve corresponds to $\chi (0)  = 0.5$.
										}
		\label{chi_vs_x}
	\end{minipage}\hfill
	\begin{minipage}[ht]{.45\linewidth}
		\begin{center}
			\fbox{
				\includegraphics[width=.9\linewidth]{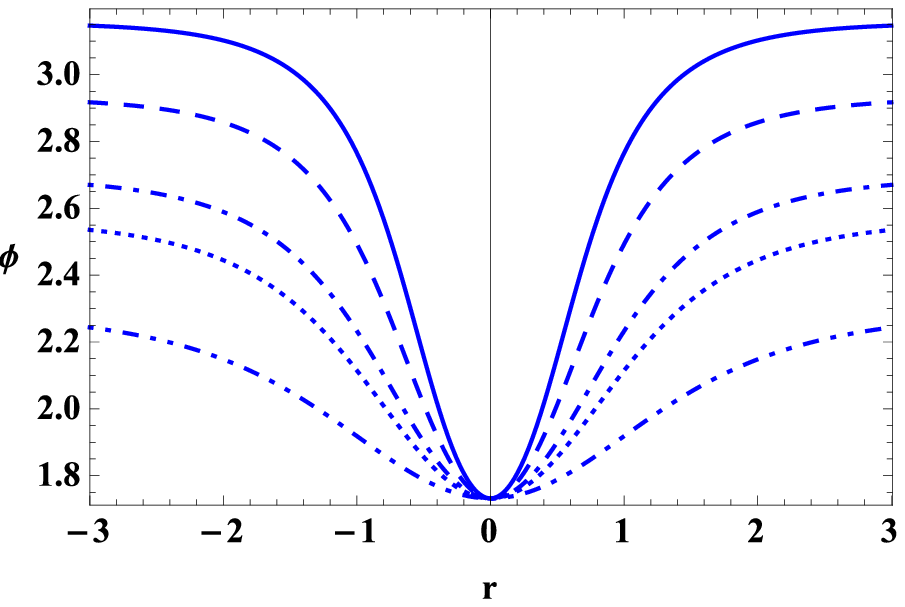}
			}
		\end{center}
		\caption{The behavior of the scalar field $\phi(r)$ for the traversable wormhole.}
		\label{phi_vs_x}
	\end{minipage}
	\begin{minipage}[ht]{.45\linewidth}
		\begin{center}
			\fbox{
				\includegraphics[width=.9\linewidth]{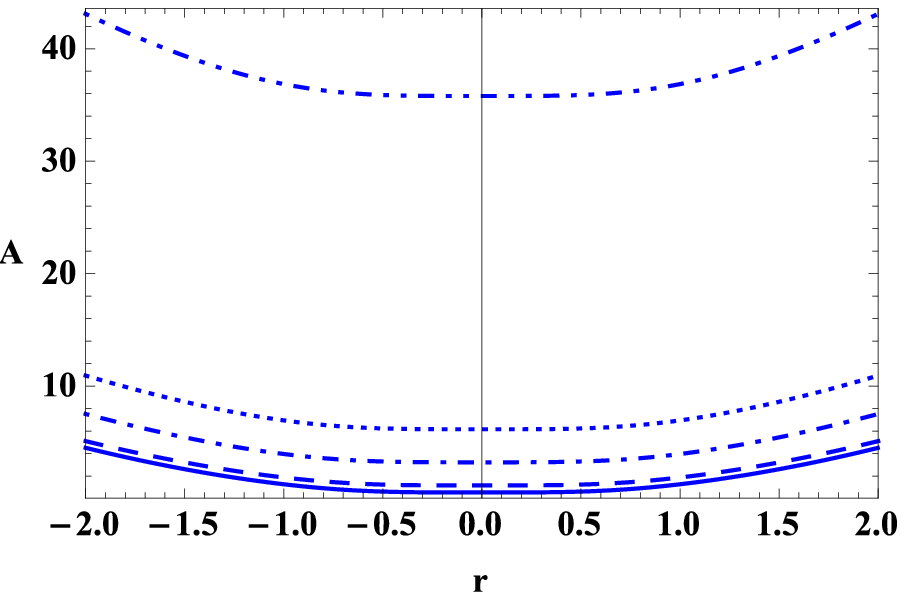}
			}
		\end{center}
		\caption{The metric function $A(r)$ for the traversable wormhole.}
		\label{A_vs_x}
	\end{minipage}\hfill
	\begin{minipage}[ht]{.45\linewidth}
		\begin{center}
			\fbox{
				\includegraphics[width=.9\linewidth]{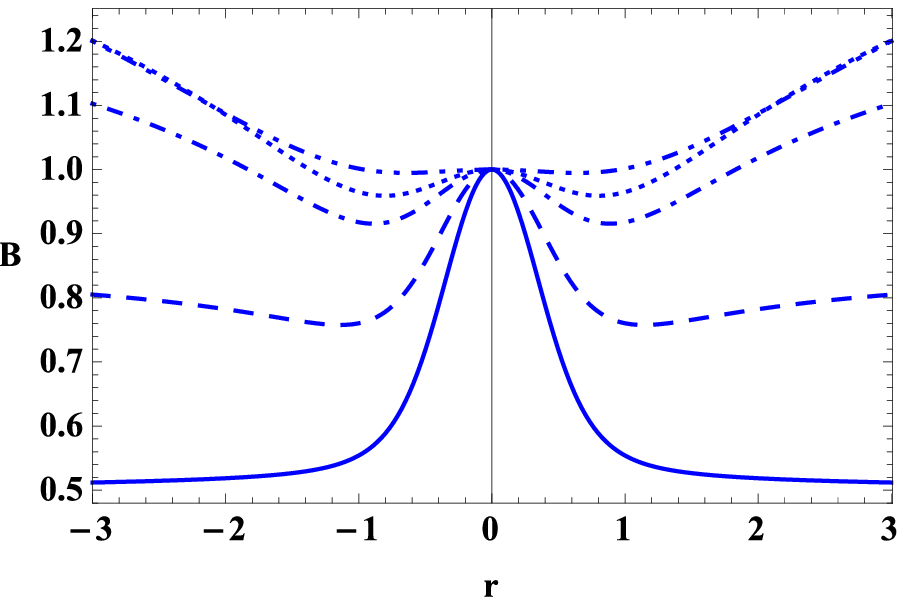}
			}
		\end{center}
		\caption{The metric function $B(r)$ for the traversable wormhole.}
		\label{B_vs_x}
	\end{minipage}
	\begin{minipage}[ht]{.45\linewidth}
		\begin{center}
			\fbox{
				\includegraphics[width=.9\linewidth]{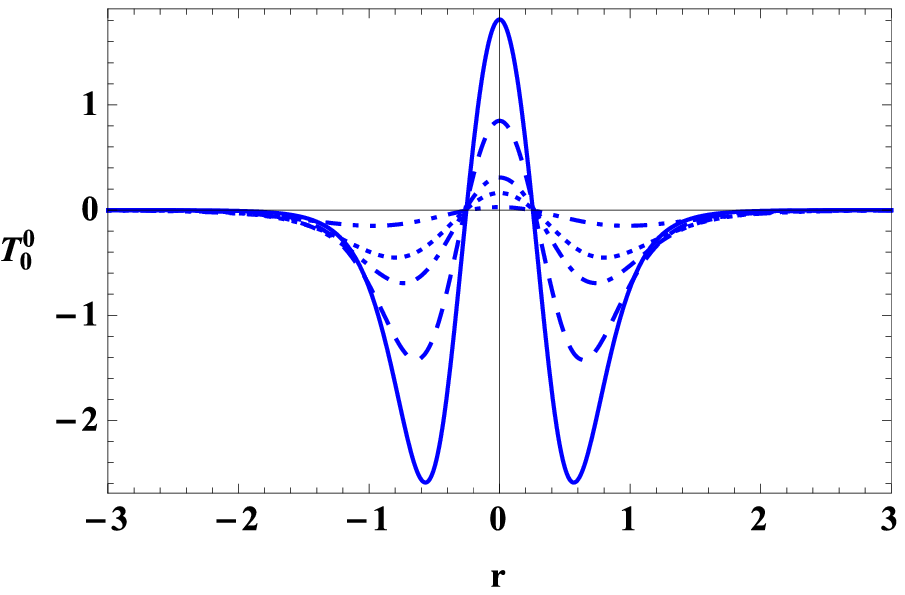}
			}
		\end{center}
		\caption{The energy densities distributions for the traversable wormhole.}
		\label{epsilon_vs_x}
	\end{minipage}\hfill
		\begin{minipage}[ht]{.45\linewidth}
		\begin{center}
			\fbox{
				\includegraphics[width=0.9\linewidth]{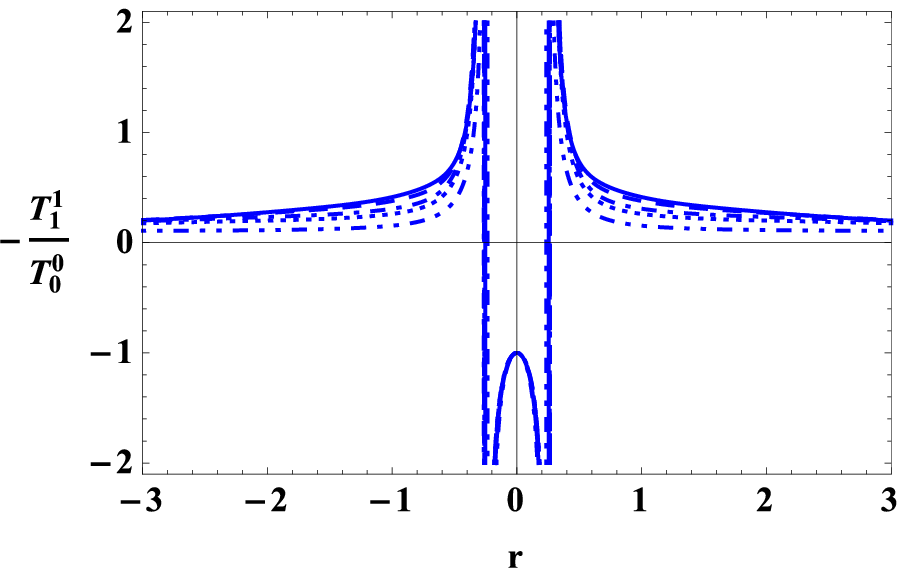}
			}
			\caption{The equation of state $w = p_{r}(r)/\epsilon(r)$ for the wormhole solutions.}
			\label{F_wh_sine2}
		\end{center}
	\end{minipage}\hfill 
	\begin{minipage}[ht]{.45\linewidth}
		\begin{center}
			\fbox{
				\includegraphics[width=.9\linewidth]{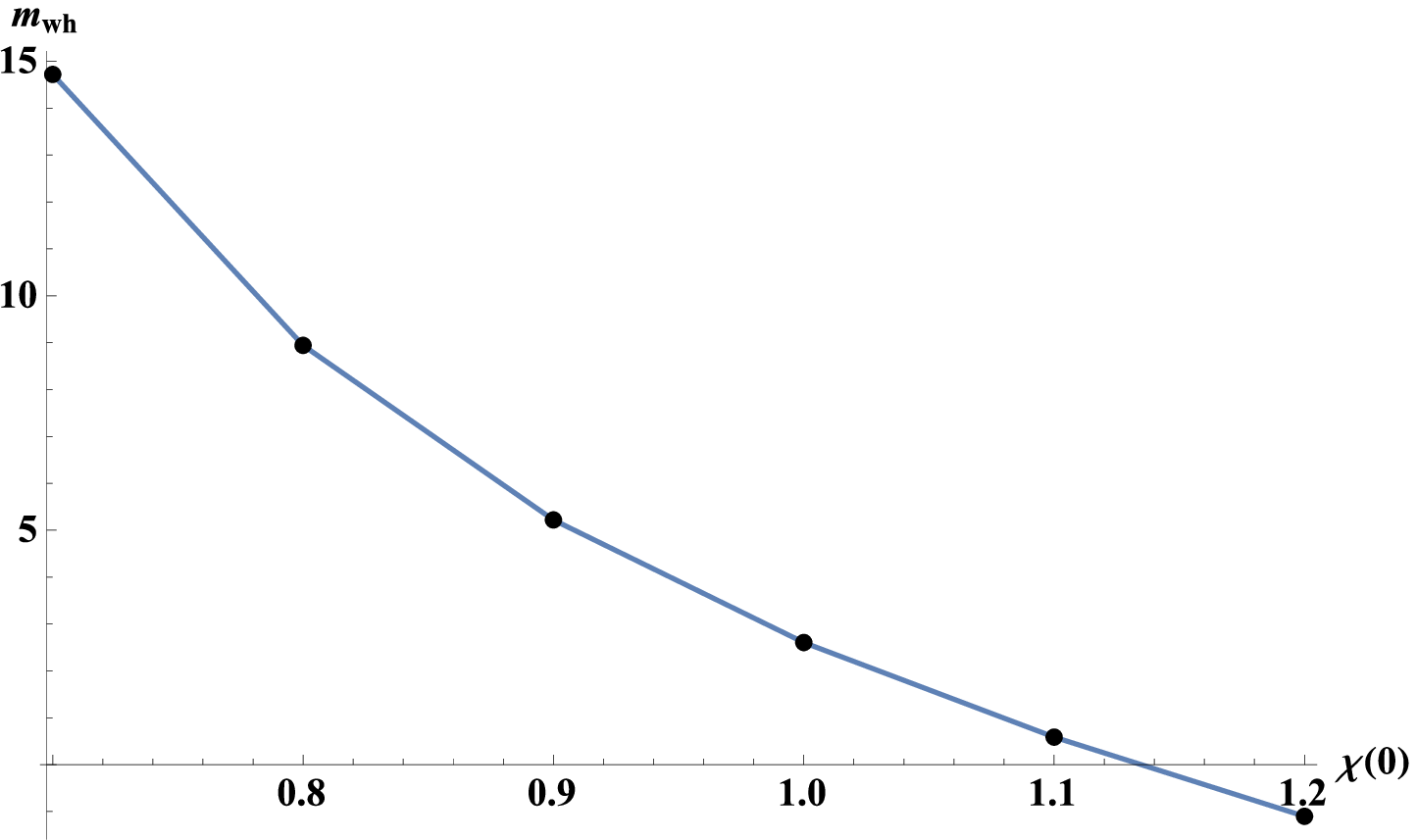}
			}
		\end{center}
	\caption{Wormhole mass as a function of $\chi (0) = 0.7, 0.8, 0.9, 1.0, 1.1, 1.2$.}
	\label{mass_vs_chi0}
	\end{minipage}
\vspace{-.5cm}
\end{figure}

We seek static solutions of Eqs.~\eqref{Einstein-gen}-\eqref{field-gen} for the wormhole metric in polar Gaussian coordinates~\cite{Visser}:
\begin{equation}
\label{metric}
  ds^2  = B(r) dt^2-dr^2-A(r)(d\theta^2+\sin^2\theta d\varphi^2),
\end{equation}
where $A(r), B(r)$ are even functions depending only on the coordinate $r$ which covers the entire range $-\infty < r < +\infty$. Using this
metric, one can obtain the following complete system of the Einstein and scalar field equations:
\begin{eqnarray}
\label{Einstein_a}
  \frac{A^{\prime \prime}}{A} - \frac{1}{2}\left(\frac{A^{\prime}}{A}\right)^2 -
  \frac{1}{2}\frac{A^{\prime}}{A}\frac{B^{\prime}}{B} &=& \phi^{\prime 2} + \chi^{\prime 2}~,
\\
\label{Einstein_b}
 \frac{A^{\prime \prime}}{A} + \frac{1}{2}\frac{A^{\prime}}{A}\frac{B^{\prime}}{B} - \frac{1}{2}\left(\frac{A^{\prime}}{A}\right)^2  - 
\frac{1}{2}\left(\frac{B^{\prime}}{B}\right)^2 + \frac{B^{\prime \prime}}{B} &=&
 2\left[\frac{1}{2}(\phi^{\prime 2} + \chi^{\prime 2}) + V\right] ,
 \\
\label{Einstein_c}
  \frac{1}{4}\left(\frac{A^{\prime}}{A}\right)^2 - \frac{1}{A} + \frac{1}{2}\frac{A^{\prime}}{A}\frac{B^{\prime}}{B} &=&
   - \frac{1}{2}(\phi^{\prime 2} + \chi^{\prime 2}) + V ,
\\
\label{field_a}
    \phi^{\prime \prime} + \left(\frac{A^\prime}{A} + \frac{1}{2}\frac{B^\prime}{B}\right)\phi^\prime =
    \phi \left[2\chi^2 + \lambda_1(\phi^2 - m_1^2)\right] ,
\\
    \label{field_b}
    \chi^{\prime \prime}+\left(\frac{A^\prime}{A}+\frac{1}{2}\frac{B^\prime}{B}\right)\chi^\prime=
    \chi \left[2\phi^2+\lambda_2(\chi^2-m_2^2)\right],
\end{eqnarray}
where the prime denotes differentiation with respect to $r$. Eq.~\eqref{Einstein_a}
is obtained by subtracting the $\left(^r _r\right)$ component of Eqs.~\eqref{Einstein-gen} from the $\left(^t _t\right)$ component,
and Eqs.~\eqref{Einstein_b} and \eqref{Einstein_c} are the $\left(^\theta _\theta\right)$ and $\left(^r _r\right)$
components of Eqs.~\eqref{Einstein-gen}. In Eqs.~\eqref{Einstein_a}-\eqref{field_b} the following rescaling has been introduced:
$r \rightarrow \sqrt{\kappa}\, r$, $\varphi \rightarrow \varphi/\sqrt{\kappa}$,
$\chi \rightarrow \chi/\sqrt{\kappa}$, $m_{1,2} \rightarrow m_{1,2}/\sqrt{\kappa}$, $A(r) \rightarrow 8 \pi G A(r)$.
The arbitrary potential constant was chosen as $V_0=(\lambda_2/4) m_2^4$ so that the value of the potential in the local minimum ({\it i.e.}
$\phi = m_1$ and $\chi =0$) was equal to zero -- $V (\phi = m_1 , \chi =0) = 0$. This choice of $V_0$ also ensures zero value of the energy
density as $r \rightarrow \pm\infty$ [see Fig.~\ref{epsilon_vs_x}].

Taking into account the $\mathbb Z_2$ symmetry of the problem, the boundary conditions are chosen in the form:
\begin{alignat}{2}
    \phi(0)      &=\sqrt{3},      & \qquad \phi^\prime(0)    &=0,
\nonumber   \\
    \chi(0)&= 0.5,0.7,0.8,1.0,1.2, 
    & \qquad \chi^\prime(0)       &=0,
\nonumber		\\
    A(0)         &=-\frac{1}{V(\phi(0), \chi(0))},   & \qquad A^\prime(0)        &=0,
\nonumber \\
    B(0)         &=1.0,          & \qquad B^\prime(0)        &=0,
\label{ini1}
\end{alignat}
where the condition for $A(0)$ is chosen so as to satisfy the constraint \eqref{Einstein_c} at $r=0$; $V(\phi(0), \chi(0))$
is the value of the potential at $r=0$; the coupling constants $\lambda_1=0.1$ and $\lambda_2=1$.
Then, using the procedure for finding solutions of Refs.~\cite{Dzhunushaliev:2007cs,Dzhunushaliev:2006xh},
we obtained the results presented in Figs.~\ref{chi_vs_x}-\ref{mass_vs_chi0}. The
values of the masses $m_{1,2}$ for the five values of $\chi(0)$ from \eqref{ini1} are given in Table \ref{eignvlsWH}.

\begin{table}[H]
\begin{center}
\begin{tabular}{ |c|c|c|c|c|c|}
  \hline
   $\chi (0)$& 0.5          & 0.7         & 0.8         & 1.0         & 1.2
\\ \hline
   $m_1$    & 2.2899327     & 2.565822825 & 2.693582264 & 2.931316954 & 3.155579948
   \\ \hline
   $m_2$    & 2.712082667 & 2.866518677 & 2.949800162 & 3.12411743   & 3.304034608
\\
  \hline
\end{tabular}
\end{center}
  \caption{The eigenvalues $m_{1,2}$ versus $\chi (0)$ with $\phi (0) =\sqrt{3}$ for phantom traversable wormholes.}
  \label{eignvlsWH}
\end{table}

The asymptotic behavior of the solutions is
\begin{eqnarray}
	A &\approx& r^2 + r_0^2 ,
\label{BC_WH_1} \\
  B &\approx& B_\infty \left( 1 - \frac{r_0^2}{r^2} \right) ,
\label{BC_WH_2} \\
	\phi &\approx &
  m_1 - C_{\phi} \frac{\exp{\left(- r \sqrt{2 \lambda_1 m_1^2}\right)}}{r} ,
\label{BC_WH_3} \\
\chi &\approx& C_{\chi}\frac{\exp{\left(- r \sqrt{ 2 m_1^2-\lambda_2 m_2^2}\right)}}{r},
\label{BC_WH_4}
\end{eqnarray}
where $r_0, B_\infty, C_{\phi}, C_{\chi}$ are integration constants.

In Fig. \eqref{F_wh_sine2} we have plotted the equation of state parameter $w=\frac{p_r}{\rho} = -\frac{T^1 _1}{T^0 _0}$
for the $r$ direction (again the equation of state for the two perpendicular directions yields 
$w=\frac{p_\theta}{\rho} = -\frac{T^2 _2}{T^0 _0} =-1$ and $w=\frac{p_\varphi}{\rho} = -\frac{T^3 _3}{T^0 _0} =-1$).
As with the phantom ball solutions there is a region near the origin where the fields have a phantom equation of state with $w<-1$. The 
divergences in $w$ again come from the places where $T^0 _0 =0$ which can be seen in Fig. \eqref{epsilon_vs_x}. Unlike the
phantom ball solutions we now show that the wormhole solutions do allow positive mass. 

The mass of the wormhole is taken to be the effective mass defined in \cite{Visser} as follows:
\begin{equation}
  m_{wh}(r)  = \frac{R(0)}{2 G} + 4 \pi \int _{R(0)} ^r T^0 _0 ~ R^2 dR 
\label{WH_mass-a} ~.
\end{equation}
The mass in \eqref{WH_mass-a} uses a wormhole metric in the form 
\begin{equation}
\label{metric-a}
  ds^2  = B(r) dt^2-dr^2-R ^2(r)(d\theta^2+\sin^2\theta d\varphi^2) ~.
\end{equation}
The ansatz function $R(r)$ in \eqref{metric-a} is related to the metric ansatz function used in
this work by $R(r) = \sqrt{A (r)}$. In the limit  $r \rightarrow \infty$ the wormhole mass, for the 
form of the metric given in \eqref{metric}, becomes 
\begin{equation}
  m_{wh} (\infty) = 4 \pi \sqrt{A(0)} + 2 \pi \int _0 ^\infty T^0 _0 \sqrt{A} \frac{dA}{dr} dr ~,
\label{WH_mass}
\end{equation}
where the energy density is given as
\begin{equation}
 T^0_0 = - \frac{{\phi^\prime}^2 + {\chi^\prime}^2}{2} -  
\frac{\lambda_1}{4}(\phi^2-m_1^2)^2-\frac{\lambda_2}{4}(\chi^4 - 2 \chi^2 m_2^2) -\phi^2 \chi^2 ~,
\label{energyDens-wh}
\end{equation}
where we set $V_0 = \frac{\lambda_2}{4} m_2 ^4$ to get the potential part energy density to have the form given in 
\eqref{energyDens-wh} The wormhole mass $m_{wh} (\infty)$ has been rescaled using the same rescaling of the masses, 
$m_{1,2}$, namely $m_{wh} (\infty) \rightarrow m_{wh} (\infty)/\sqrt{8 \pi G}$. The profile of $m_{wh}$ as a 
function of $\chi (0)$ is shown in Fig.~\ref{mass_vs_chi0}.

\section{Phantom cosmic strings}
\label{pcs}

In this section we consider the case of an extended one dimensional solution -- the global cosmic string. Global cosmic strings
are built from scalar fields only. In contrast local cosmic string involve scalar fields plus gauge fields. 
This type of solution has cylindrical symmetry and we describe it via the following metric
\begin{equation}
\label{metric_string}
  ds^2 = e^{2 \nu(\rho)} dt^2 - e^{2 (\gamma(\rho) - \psi(\rho))} d\rho^2 -
  e^{2 \psi(\rho)} dz^2 - \rho^2 e^{-2 \psi(\rho)} d\varphi^2 ,
\end{equation}
and the assumption that the scalar fields are only dependent on $\rho$.  Substituting this into the
Einstein and scalar field equations \eqref{Einstein-gen}-\eqref{field-gen}, one can obtain the following set of equations:
\begin{eqnarray}
  \frac{\gamma^\prime}{\rho} - {\psi^\prime}^2 &=&
  - \kappa \left(
    \frac{1}{2} {\phi^\prime}^2 + \frac{1}{2} {\chi^\prime}^2 +
    e^{2 (\gamma - \psi)} V(\phi, \chi)
  \right) ,
\label{ef_1} \\
  \frac{\nu^\prime + \psi^\prime}{\rho} - {\psi^\prime}^2 &=&
  - \kappa \left(
    \frac{1}{2} {\phi^\prime}^2 + \frac{1}{2} {\chi^\prime}^2 -
    e^{2 (\gamma - \psi)} V(\phi, \chi)
  \right) ,
\label{ef_2} \\
  \psi^{\prime \prime } - \nu^{\prime \prime} - \psi^\prime \gamma^\prime +
  \nu^\prime \gamma^\prime - {\nu^\prime}^2 +
  \frac{\psi^\prime + \gamma^\prime - \nu^\prime}{\rho} &=&
  \kappa \left(
    - \frac{1}{2} {\phi^\prime}^2 - \frac{1}{2} {\chi^\prime}^2 -
    e^{2 (\gamma - \psi)} V(\phi, \chi)
  \right) ,
\label{ef_3} \\
  - \psi^{\prime \prime } - \nu^{\prime \prime} + \psi^\prime \gamma^\prime +
  \nu^\prime \gamma^\prime - 2 {\psi^\prime}^2 - 2\psi^\prime \nu^\prime -
  {\nu^\prime}^2 &=& \kappa \left(
    - \frac{1}{2} {\phi^\prime}^2 - \frac{1}{2} {\chi^\prime}^2 -
    e^{2 (\gamma - \psi)} V(\phi, \chi)
  \right) ,
\label{ef_4} \\
    \phi^{\prime \prime} + \phi^\prime \left(
      \frac{1}{\rho} - \gamma^\prime + \psi^\prime + \nu^\prime
    \right) &=& e^{2 (\gamma - \psi)} \phi \left[
      2 \chi^2 + \lambda_1 \left( \phi^2 - m_1^2 \right)
    \right] ,
\label{ef_5} \\
    \chi^{\prime \prime} + \chi^\prime \left(
      \frac{1}{\rho} - \gamma^\prime + \psi^\prime + \nu^\prime
    \right) &=& e^{2 (\gamma - \psi)} \chi \left[
      2 \phi^2 + \lambda_2 \left( \chi^2 - m_2^2 \right)
    \right] .
\label{ef_6} 
\end{eqnarray}
As in the case of the phantom balls and the traversable wormhole we have taken the potential constant as
$V_0 = (\lambda _2/4) m_2 ^4$ so that the potential will go to zero as $\rho \rightarrow \infty$.

To simplify the above equations we also make the additional assumption that
two of the metric functions are equal {\it i.e.} $\nu = \psi$. After some algebraic manipulations, and performing the rescaling
$\rho/\sqrt \kappa \rightarrow \rho$, $\phi \sqrt \kappa \rightarrow \phi$, $\chi \sqrt \kappa \rightarrow \chi$,
and $m_{1,2} \sqrt \kappa \rightarrow m_{1,2} $ we get the following equations for the metric functions
$\gamma(\rho), \psi(\rho)$ and phantom scalar fields $\phi(\rho), \chi(\rho)$:
\begin{eqnarray}
  \frac{\gamma^\prime}{\rho} - {\psi^\prime}^2 &=&
  - \left(
    \frac{1}{2} {\phi^\prime}^2 + \frac{1}{2} {\chi^\prime}^2 +
    e^{2 (\gamma - \psi)} V(\phi, \chi)
  \right) ,
\label{cosmic_string_1} \\
  2 \frac{\psi^\prime}{\rho} - {\psi^\prime}^2 &=&
  - \left(
    \frac{1}{2} {\phi^\prime}^2 + \frac{1}{2} {\chi^\prime}^2 -
    e^{2 (\gamma - \psi)} V(\phi, \chi)
  \right) ,
\label{cosmic_string_1a} \\
  \psi^{\prime \prime } + \frac{\psi^\prime}{\rho} &=&
  e^{2 (\gamma - \psi)} \left(
    1 - 2 \rho \psi^\prime
  \right) V(\phi, \chi) ,
\label{cosmic_string_2} \\
    \phi^{\prime \prime} + \phi^\prime \left(
      \frac{1}{\rho} - \gamma^\prime + 2 \psi^\prime
    \right) &=& e^{2 (\gamma - \psi)} \phi \left[
      2 \chi^2 + \lambda_1 \left( \phi^2 - m_1^2 \right)
    \right] ,
\label{cosmic_stringl_3} \\
    \chi^{\prime \prime} + \chi^\prime \left(
      \frac{1}{\rho} - \gamma^\prime + 2 \psi^\prime
    \right) &=& e^{2 (\gamma - \psi)} \chi \left[
      2 \phi^2 + \lambda_2 \left( \chi^2 - m_2^2 \right)
    \right] ,
\label{cosmic_string_4} 
\end{eqnarray}
where the prime denotes differentiation with respect to the rescaled radial coordinate $\rho$.

\begin{figure}[H]
	\begin{minipage}[ht]{.45\linewidth}
		\begin{center}
			\fbox{
				\includegraphics[width=.9\linewidth]{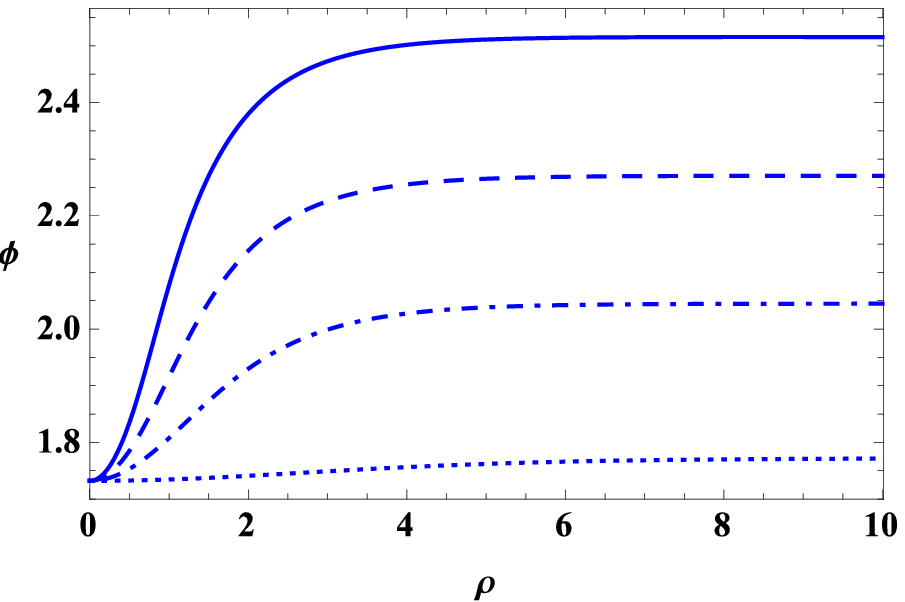}
			}
		\end{center}
		\caption{The scalar fields $\phi(\rho)$ for the phantom cosmic strings.
                   For Figs. \ref{phi_vs_x_string} -- \ref{epsilon_vs_x_string} 
                    the solid curve corresponds to $\chi (0)  = 0.8$,
                    the dashed curve  corresponds to $\chi (0)  = 0.6$,
                    the dashed - dotted corresponds to $\chi (0)  = 0.4$,
                    the dotted curve corresponds to $\chi (0)  = 0.1$}
	\label{phi_vs_x_string}
	\end{minipage}\hfill
	\begin{minipage}[ht]{.45\linewidth}
		\begin{center}
			\fbox{
				\includegraphics[width=.9\linewidth]{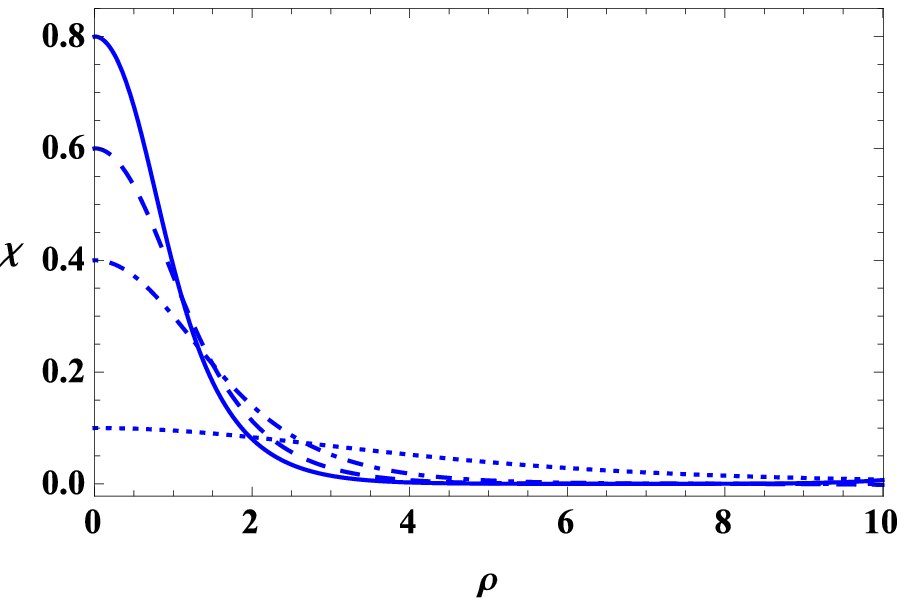}
			}
		\end{center}
		\caption{The scalar fields $\chi(\rho)$ for the phantom cosmic strings.
										}
	\label{chi_vs_x_string}
	\end{minipage}
	\begin{minipage}[ht]{.45\linewidth}
		\begin{center}
			\fbox{
				\includegraphics[width=.9\linewidth]{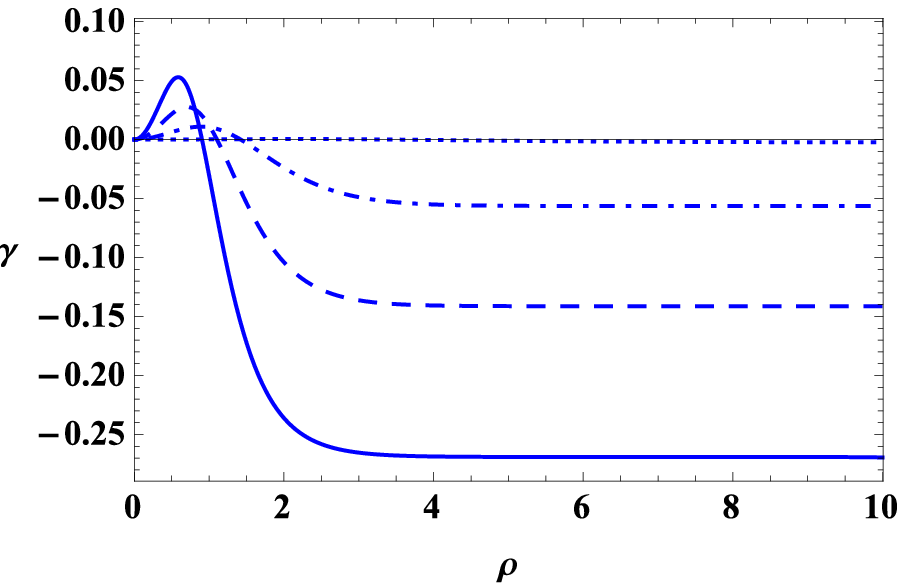}
			}
		\end{center}
		\caption{The metric functions $\gamma(\rho)$ for the phantom cosmic strings.
										}
	\label{A_vs_x_string}
	\end{minipage}\hfill
	\begin{minipage}[ht]{.45\linewidth}
		\begin{center}
			\fbox{
				\includegraphics[width=.9\linewidth]{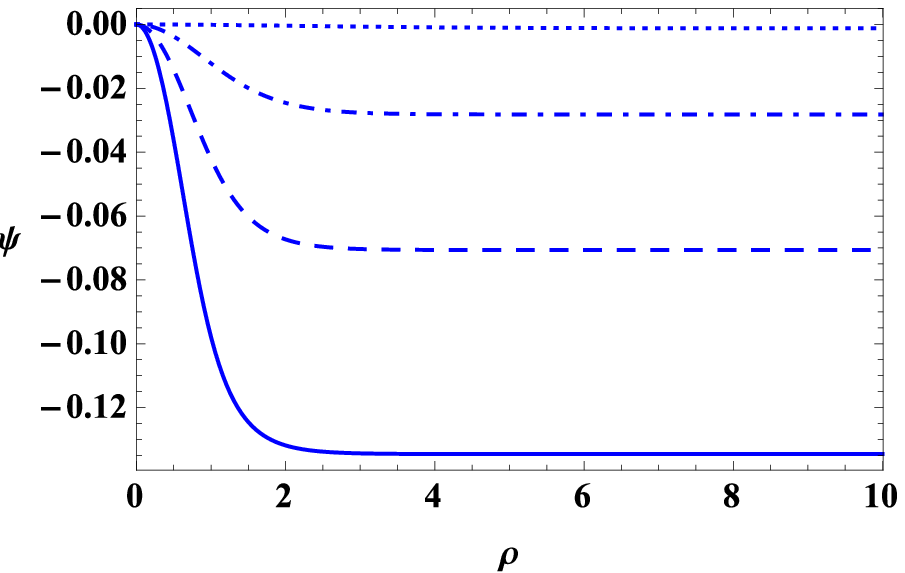}
			}
		\end{center}
		\caption{The metric functions $\psi(\rho)$ for the phantom cosmic strings.
											}
	\label{B_vs_x_string}
	\end{minipage}
	\begin{minipage}[ht]{.45\linewidth}
		\begin{center}
			\fbox{
				\includegraphics[width=.9\linewidth]{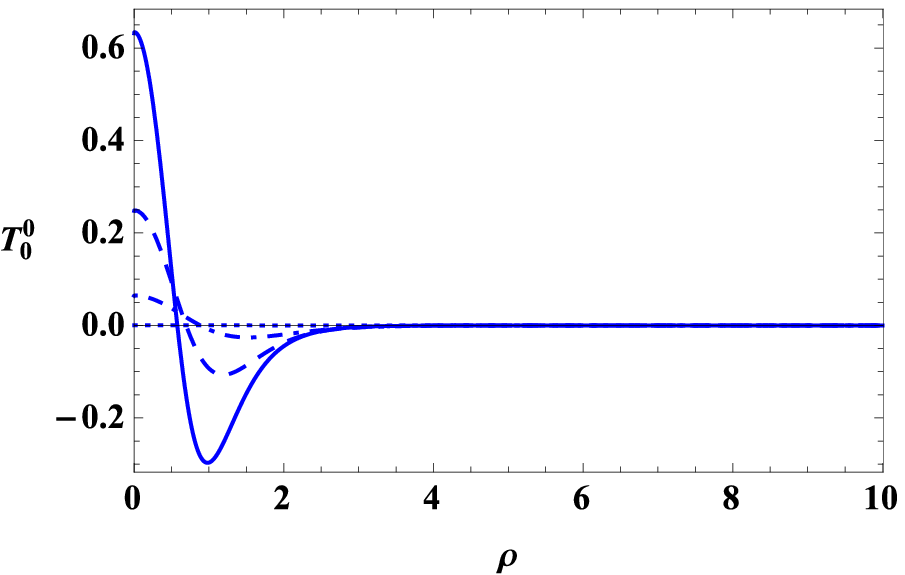}
			}
		\end{center}
		\caption{The energy density profiles for the phantom cosmic strings.
										}
	\label{epsilon_vs_x_string}
	\end{minipage}\hfill
	\begin{minipage}[ht]{.45\linewidth}
			\begin{center}
			\fbox{
				\includegraphics[width=.9\linewidth]{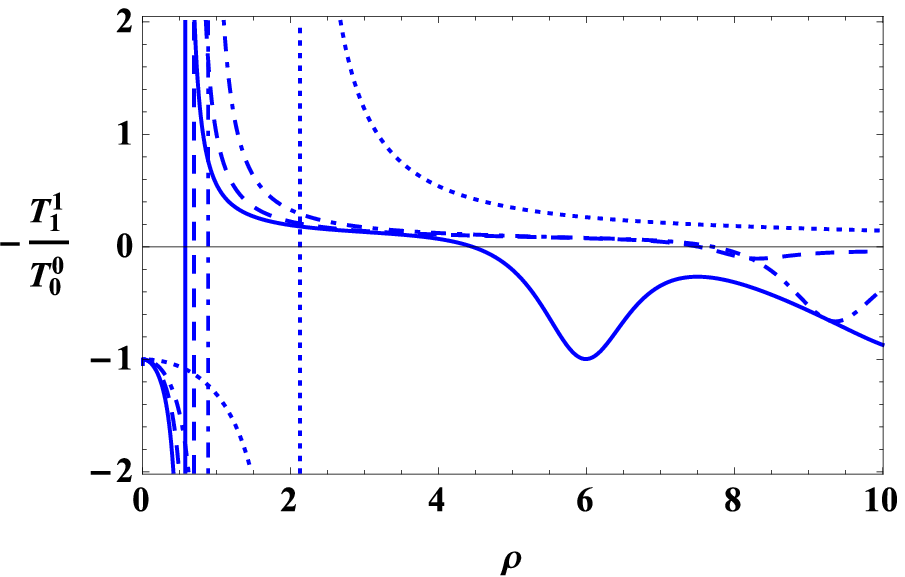}
			}
		\end{center}
		\caption{The equation of state $w = p_{\rho}(\rho)/\epsilon(\rho)$ for the cosmic string solutions.
										}
	\label{w_string}
	\end{minipage}\hfill
	\end{figure}

To numerically integrate the above equations we will use \eqref{cosmic_string_1}, \eqref{cosmic_string_2},
\eqref{cosmic_stringl_3}, and \eqref{cosmic_string_4}, along with the definition of $V(\phi, \chi)$
in \eqref{pot}. Equation \eqref{cosmic_string_1a} is redundant.
The boundary conditions are chosen at the center ($\rho=0$) in the form:
\begin{alignat}{2}
    \label{ini1_CS}
    \gamma(0)& = 0,       & \qquad 
\nonumber \\
    \psi(0)& = 0,       & \qquad \psi'(0)& = 0 ,
  \\
    \phi(0)&= \sqrt 3 ,   & \qquad \phi^\prime(0)       &=0,
	\nonumber \\
    \chi(0)&= 0.1, 0.4, 0.6, 0.8 ,   & \qquad \chi^\prime(0)       &=0.		\nonumber
\end{alignat}
As before, for a given initial value of $\chi(0)$, there are special values (eigenvalues) for $m_1, m_2$ for which
solutions are found with acceptable asymptotic behavior. The procedure for finding
these special values of $m_1, m_2$ is that given in Refs.~\cite{Dzhunushaliev:2007cs,Dzhunushaliev:2006xh}.
The value of these masses $m_1, m_2$ as a function of $\chi(0)$ is shown in Table \ref{eignvlsCS}.
\begin{table}[H]
\begin{center}
\begin{tabular}{ |c|c|c|c|c|}
  \hline
   $\chi (0)$ & 0.1             & 0.4             & 0.6             & 0.8             \\ \hline
   $m_1$     & 1.7720039 & 2.044828   & 2.2706674 & 2.515715    \\ \hline
   $m_2$     & 2.4896351 & 2.7060905 & 2.8649473 & 3.0277392   \\
  \hline
\end{tabular}
\end{center}
  \caption{The eigenvalues $m_{1,2}$ versus $\chi (0)$ with $\phi (0) = 0.5$ for phantom cosmic strings. 
	The coupling constants $\lambda_1=0.1, \lambda_2=1$.}
  \label{eignvlsCS}
\end{table}
The results of numerical calculations for the scalar fields, $\phi$ and $\chi$, are given in Figs.~\ref{phi_vs_x_string},
\ref{chi_vs_x_string}. The behavior of the scalar fields is similar to what was found for the phantom balls.
The scalar field $\phi$ started at $\phi =0$ and as $\rho \rightarrow \infty$ approached some constant, non-zero value. The
scalar field $\chi$ started at some non-zero value and approached $\chi =0$ as $\rho \rightarrow \infty$. The metric functions
$\gamma(\rho), \psi(\rho)$ are given in Figs.~\ref{A_vs_x_string},
\ref{B_vs_x_string}. These metric functions both approach some constant negative value as $\rho \rightarrow \infty$.
The energy density, $T_0^0$, is given in Fig.~\ref{epsilon_vs_x_string}. Near $\rho \approx 0.5$ the energy density, $T_0^0$, 
changes sign and then goes asymptotically to zero from below.
	
Since the cosmic string is of infinite length its total mass is infinite, but the relevant quantity is mass per unit length. To obtain this
we integrate the energy density, $T_0 ^0$, over the full range of $\rho$ and $\varphi$ but along $z$ we integrate a finite length, $L$, and 
then divide by this to get mass per unit length. The energy density \eqref{emt} is
\begin{equation}
\label{emt-cs}
T_0 ^0 = - \frac{1}{2}({\phi '}^2 + {\chi '}^2 ) e^{(2\psi - 2\gamma)} - V(\phi, \chi) ~.
\end{equation}
Now the mass of a length $L$ of the cosmic string is
\begin{equation}
\label{cs-mass}
m_{cs} = 2 \pi L \int _0 ^\infty T^0 _0 \rho e^{(\gamma - \psi)} d \rho ~,
\end{equation}
where the $\varphi$ integration gives $2 \pi$ and the $z$ integration gives $L$. The factor $\rho e^{(\gamma - \psi)}$ comes from 
$\sqrt{- det ~ g ^3}$, the volume factor for the spatial part of the metric in \eqref{metric_string}. From \eqref{cs-mass}
and \eqref{emt-cs} the mass per unit length, $\mu _{cs} \equiv m_{cs}/ L$, of the cosmic string solution is
\begin{equation}
\label{mu-cs}
\mu _{cs} = 2 \pi \int _0 ^\infty T_0 ^0 \rho e^{(\gamma - \psi)} d \rho = \int _0 ^\infty \left[ - \frac{\rho}{2} 
\left( {\phi '}^2 + {\chi '}^2 \right) e^{(\psi - \gamma)} - \rho V(\phi , \chi ) e^{(\gamma - \psi)} \right] d \rho ~.
\end{equation} 
We calculated $\mu _{cs}$ for a range of different initial values of $\chi (0)$ and $\phi (0)$ and in general found that
$\mu _{cs} < 0$. In one case when $\phi (0) =0$ and $\chi (0) = 0$ we did find by extrapolation that a very small positive value
for $\mu _{cs}$. However, for this point we could not trust the extrapolation. Thus in this case the conclusion was
that $\mu _{cs} < 0$ in general.     

The equation of state parameter, $w = -\frac{T^1_1}{T^0_0}$ for the cosmic string is shown in Fig. \eqref{w_string}. As in the
previous three cases we have defined the equation of state parameter in terms of the radial pressure 
(for the cosmic string this means pressure in the $\rho$-direction). The equation of state parameter has some range of $\rho$ near
$\rho =0$ for which $w<-1$ ({\it i.e.} the scalar fields have a phantom equation of state) and
so we call these solutions phantom cosmic strings. As with the wormhole and phantom ball solutions
the region where one has phantom behavior is near the origin. From Fig. \eqref{w_string} one can see that $w$ diverges. These
divergences occur at the places where $T^0 _0 \rightarrow 0$ as shown in Fig. \eqref{epsilon_vs_x_string}.

\section{Discussion and conclusions}

We have studied several hypothetical compact and extended astrophysical objects supported by two phantom scalar fields.
When distributed homogeneously and isotropically over the Universe such phantom fields could be the source of the 
present accelerated expansion of the Universe. However, one can imagine situations when in the process of evolution of the
Universe there could arise inhomogeneous distributions of such phantom fields that might lead to the formation of localized or
partially localized configurations such as those considered here.

The ``phantom" domain walls investigated in section \eqref{DW} are in fact not phantom since, despite having
negative sign kinetic energy terms for the scalar fields, the equation of state parameter associated with
the $x$ direction is always $w \ge -1$  as shown in Fig. \eqref{F_wh_sine}. Leaving aside the equation of state
parameter, these domain wall solutions do have the ``unphysical" characteristic that their
energy density, $T^0 _0$, is everywhere negative $T^0 _0 <0$. This leads one to strongly suspect that these solutions 
are unstable {\it i.e.} that they would dissipate rapidly. This is actually a good, physical feature since normal domain walls
tend to dominate the energy density of the Universe \cite{zeldovich} and can be ruled out observationally.
Thus the domain walls studied here, exactly because of their hypothesized unstable nature, could
be of more consequence than normal domain walls -- these domain walls might form and quickly dissipate, perhaps leaving
some small imprint on the CMB. In regard to observational constraints regular domain walls are more problematic. However,
one would still need to confirm that these domain walls are unstable by performing some stability analysis as in \cite{dzh-folo-2010},
and one would need to determine if they would leave some imprint on observables from the early Universe.

In section \eqref{SSS} we found phantom ball solutions. In contrast to the domain wall solutions, the phantom ball
solutions can properly be call ``phantom solutions" since for some range of $r$ near the origin their equation of
state parameter does satisfy $w< -1$ as seen in Fig. \eqref{F_wh_sine1}. These solutions might be viewed as phantom versions of 
magnetic monopole solutions \cite{thooft} that occur in Grand Unified Theories (GUT), except that the phantom ball
solutions of section \eqref{SSS} involve scalar fields coupled to gravity, while the GUT monopoles are a combination of gauge 
and scalar fields. These GUT monopoles have been shown to
be problematic to cosmological observations \cite{preskill} (one can solve the ``GUT monopole problem" via
inflation). The phantom ball solutions had some range of $r$ where the energy density was positive, $T^0 _0 >0$ as
seen in Fig. \eqref{epsilon_vs_x_boson_stars}. However, when integrating the energy density over all space we found that the 
mass of the phantom balls was negative as seen in Fig. \eqref{mass_vs_chi0_boson_stars}. Thus as with the domain walls, the phantom ball
solutions are probably unstable (of course for both solutions one should check this in detail, but it would be odd and interesting 
if either of these solutions were stable under a more detailed analysis). As with the domain walls, this indication
that the phantom ball solutions are unstable/unphysical is good from the physical standpoint in that both of these
solutions would not remain around long due to their postulated instability. However, these two solutions might form
in the early Universe and leave some signature on observables from the early Universe
({\it e.g.} they might leave an small imprint on the CMB) yet due to their postulated instability, decay away before causing
problems with cosmological observations like those associated with regular domain walls and/or GUT magnetic monopoles.     

In section \eqref{pwh} we found wormhole solutions supported by the phantom fields. The wormhole solutions were the only 
solutions of the four types that we studied that had (at least for some set of parameters) positive mass as shown by Fig. 
\eqref{mass_vs_chi0}. Since these solutions did not have any horizons ({\it i.e} the metric ansatz functions, $A(r)$ and $B(r)$,
did not got to zero or become negative) the wormholes are traversable ``in principle" meaning \cite{Visser} simply that
they do not have horizons. In order to determine if these wormhole solutions are travesable ``in practice" (meaning that
the tidal forces can be engineered to be small enough to allow an observer of some given length, $l_0$, and able to withstand
some local gravitational acceleration of $g$, to pass through the wormhole throat without being tidally disrupted \cite{Visser}) is more
difficult to determine, especially given that the solutions we have obtained are numerical. However, given the wide range of 
shapes for the metric ansatz functions $A(r)$ and $B(r)$, it is likely that one could also engineer these phantom wormholes to be
traversable ``in practice". 

Finally in section \eqref{pcs} we found global, phantom cosmic string solutions. Similar to the cases of the phantom ball
solutions and phantom wormhole solutions, the equation of state parameter satisfied $w<-1$ for some range of $\rho$,
as shown in Fig. \eqref{w_string}, making the term phantom cosmic string appropriate. Also, as in the case of the phantom ball solutions,
although for some ranges of $\rho$ the energy density of these cosmic strings was positive, $T^0 _0 >0$, 
the mass per unit length was negative, $\mu _{cs} < 0$. (There was some very particular choice of initial conditions, 
namely $\phi (0) =0$ and $\chi (0) = 0$, which might have given a very small, positive $\mu _{cs}$, but we could 
not determine if this was a numerical artifact or not). As with the domain wall solutions and the phantom ball solutions
the fact that $\mu _{cs} < 0$ indicates that these solutions are likely not stable. Now in the case of the domain wall
and phantom ball solutions this postulated instability is desirable from a physical point of view, since ordinary
domain walls and GUT magnetic monopole solutions (the rough equivalent of the phantom ball solutions) cause 
difficulties in regard to cosmological observations. 
On the other hand normal cosmic strings are thought to play a potential role in structure formation 
\cite{Vilen} in the Universe. The postulated instability of our phantom cosmic string solution would rule out 
their use for structure formation, but as with the domain wall and phantom ball solutions, the phantom cosmic 
strings might leave some imprint on cosmological observables such as the CMB.

\section*{Acknowledgments}

This work was supported by a grant $\Phi.0755$  in fundamental research in natural sciences by the
Ministry of Education and Science of Kazakhstan. The work of DS is supported by a 2015-2016 Fulbright Scholars Grant to Brazil.
DS wishes to thank the ICTP-SAIFR in S{\~a}o Paulo for its hospitality. \\

{\bf Note added:} After the submission of the article proofs we learned of a review article \cite{cai} and three other articles
\cite{sari} which discuss aspects of phantom fields relevant to the study in this work.

\end{document}